# Distinguishing Electronic Band Structure of Single-layer and Bilayer Ruddlesden-Popper Nickelates Probed by *in-situ* High Pressure X-ray Absorption Near-edge Spectroscopy

Mingtao Li[1,*,#], Yiming Wang[1,*], Cuiying Pei[2,*], Mingxin Zhang[2,*], Nana Li[1], Jiayi Guan[1], Monica Amboage[3], N-Diaye Adama[4], Qingyu Kong[4], Yanpeng Qi[2,5,6,#], and Wenge Yang[1,#]

[1]Center for High Pressure Science and Technology Advanced Research, Shanghai 201203, China

[2]School of Physical Science and Technology, ShanghaiTech University, Shanghai 201210, China

[3]Diamond Light Source Ltd, Diamond House, Harwell Science and Innovation Campus, Didcot, Oxfordshire OX11 0DE, UK

[4]Synchrotron SOLEIL, L'Orme des Merisiers, Saint-Aubin, BP48, 91192 Gif-sur-Yvette, France

[5]ShanghaiTech Laboratory for Topological Physics, ShanghaiTech University, Shanghai 201210, China

[6]Shanghai Key Laboratory of High-resolution Electron Microscopy, ShanghaiTech University, Shanghai 201210, China

[*] These authors contributed equally to this work.

[#]Corresponding authors:

mingtao.li@hpstar.ac.cn (M.T.L.), qiyp@shanghaitech.edu.cn (Y.P.Q.), yangwg@hpstar.ac.cn (W.G.Y.)

**ABSTRACT**

Ruddlesden-Popper (RP) phase $La_{n+1}Ni_nO_{3n+1}$ ($n$=2) was recently discovered to host superconductivity (SC) with critical temperature $T_c$~80 K under pressures. Theoretical calculations show that Ni $3d_{3z^2-r^2}$ and $3d_{x^2-y^2}$ orbitals dominate near Fermi energy ($E_F$), and the interlayer interaction mediated by inner apical oxygen atoms induces the splitting of $d_{3z^2-r^2}$ derived band into antibonding and bonding bands, which are believed to be essential for the emerged SC. To date, no experimental electronic band structure results are yet reported to validate the related theoretical proposals at high pressures (HP). Here, we report a comprehensive study of electronic band structure for single-layer (SL) and bilayer (BL) RP-nickelates probed by *in-situ* HP X-ray absorption near edge spectroscopy (XANES). At ambient pressure (AP), the energy splitting $\Delta E$ of $d_{3z^2-r^2}$ and $d_{x^2-y^2}$ bands are directly observed in La$_3$Ni$_2$O$_7$ (BL-La327) but not in La$_2$NiO$_4$ (SL-La214) above $E_F$ from both experimental and theoretical simulated XANES, underlining the critical role of inner apical O atoms. A combination of DFT-based electronic band structure and projected density of states (PDOS) calculations with simulated XANES enables us to explain the observed main XANES features labelled by $a$, $A$, $B'/B$ and $C$ when considering the orbital hybridizations, crystal field splitting (CFS) and core-hole screening of different $3d$ configurations for SL-La214 and BL-La327 nickelates. At HP, the $3d$ orbital $\Delta E$ values form a dome-like evolution for $P \geq 7.7$ GPa with the maximum locating at around 20 GPa for metallic BL-La327. In contrast, both peak positions of $d_{3z^2-r^2}$ and $d_{x^2-y^2}$ bands shift downwards to $E_F$ for the metallic BL-La327 with strong superconducting transition at $T_c \sim 80$ K for $P \geq 20$ GPa, while the peak of dominant antibonding $d_{3z^2-r^2}$ band gradually lifts away from $E_F$ for the semiconducting BL-La327. Analysis of integrated area and full width at half maximum (FWHM) of $a$ peak provides strong evidence that the bonding $d_{3z^2-r^2}$ band crosses $E_F$ above about 7.7 GPa forming a hole-like pocket to increase hole counts up to about 20 GPa for the metallic BL-La327. Growth of integrated area of $a$ and $C$ peaks further evidences pressure-induced hole doping effect. Meanwhile, the pressure dependent FWHM of $a$ peak implies a nonmonotonic ($\propto$-shape like) evolution of orbital-selective electronic correlation for $P \geq 7.7$ GPa with extrema emerging at about 20 GPa. Moreover, we demonstrate that the energy shift of main edges is mainly contributed by the pressure-induced lattice contraction instead of valence change of Ni ions according to Natoli's rule. Alternatively, we estimate the relative hole doping level using the energy shift of $a$ peak, yielding 0.074 hole per Ni site or equivalently $1.1 \times 10^{21}$ cm$^{-3}$ at 20 GPa for the metallic BL-La327, which is comparable to high-$T_c$ cuprates. Our results have timely examined the electronic band structures as obtained from theoretical calculations, emphasizing the essential role of both $d_{3z^2-r^2}$ and $d_{x^2-y^2}$ bands as well as the electronic correlation in superconducting pairing for pressurized La$_3$Ni$_2$O$_7$.

I. INTRODUCTION

The discovery of high-$T_c$ SC in layered (La,Ba)$_2$CuO$_4$ ($T_c\sim$35 K) cuprates have generated enormous interests in condensed matter physics [1]. As an isostructural compound to the K$_2$NiF$_4$-type La$_2$CuO$_4$, the SL *RP*-phase La$_2$NiO$_4$ (SL-La214) are well studied to explore the possible SC by hole doping through chemical substitution of Sr for La [2]. However, no SC is observed in bulk La$_{2-x}$Sr$_x$NiO$_4$ despite its metallization induced by high Sr-doping level at $x$=1.3 [2]. This is in sharp contrast to the (La$_{1-x}$Sr$_x$)$_2$CuO$_4$ high-$T_c$ cuprates [3], doping level $x$=0.01 readily induces metallicity and bulk SC sets in at $x$=0.03. Without hole doping, SL-La214 is an antiferromagnetic (AFM) charge transfer (CT) insulator with spin $S$=1 for formally Ni$^{2+}$ ($3d^8$) [4,5], while La$_2$CuO$_4$ is an AFM insulator with spin $S$=1/2 for formally Cu$^{2+}$ (3d$^9$) [6,7].

On the other hand, the Ni$^{1+}$ ($3d^9$) (for instance, infinite layer LaNiO$_2$ without apical O atoms) have been theoretically discussed to mimic the $3d^9$ configuration of Cu$^{2+}$ of high-$T_c$ cuprates. Lately, findings of SC at 9-15 K in Sr-doped infinite layered (IL) Nd$_{0.8}$Sr$_{0.2}$NiO$_2$ single-crystal thin films have renewed the research upsurge in achieving unconventional high-$T_c$ nickelates [8]. By applying pressure, $T_c$ can be enhanced to about 31 K at 12.1 GPa in superconducting Pr$_{0.82}$Sr$_{0.18}$NiO$_2$ thin films [9]. In parent IL-nickelates, the Ni $3d_{x^2-y^2}$ orbital is self-doped by a small number of holes due to the presence of electron pockets originating from the La $5d$ orbitals [10,11]. Compared to cuprates, the essential role of $d_{x^2-y^2}$ band is suggested to understand its lower $T_c$ due to the strong self-energy renormalized effect [10]. Besides, a larger $p-d$ energy splitting, denoted as CT gap $\Delta_{CT}$, is revealed in IL-nickelates compared to CaCuO$_2$ cuprate counterpart [11]. Importantly, IL-nickelates have the similarly large value of the longer-range in-plane hopping $t'$ and the energy splitting ($\Delta E$) of $d_{x^2-y^2}$ and $d_{3z^2-r^2}$ bands compared to those of cuprates [11]. The factors of $\Delta_{CT}$ magnitude [12], the strength of $t'/t$ ratio with $t$ being nearest-neighbor hopping [13], and $\Delta E$ magnitude [14], have been proposed to be essential for understanding SC in cuprates.

Very recently, Sun et al reported signatures of pressure-induced SC at 80 K in the BL RP-phase La$_{n+1}$Ni$_n$O$_{3n+1}$ ($n$=2, denoted as BL-La327) for 14.0 GPa≤$P$≤43.5 GPa [15]. Following studies have confirmed the high-$T_c$ SC in BL-La327 [16-18], while no SC is observed up to 50 GPa for SL-La214 [16]. At AP, SL-La214 crystallizes into a tetragonal structure ($I4/mmm$, space group No. 139, $Z$=2) with elongated NiO$_6$ octahedrons [2,19], as seen in Fig. 1(a). BL-La327 forms a orthorhombic structure ($Amam$, space group No. 63, $Z$=4) as displayed in Fig. 1(b) [20], which possesses a formal Ni$^{2.5+}$ valence state with $3d^{7.5}$ configuration assuming the fully ionic La$^{3+}$ and O$^{2-}$ states. Here, we denote the inner apical oxygen and outer apical oxygen as O$_{iap}$ and O$_{oap}$, respectively. Specially, Ni$^{2.5+}$ is suggested to be the mixed-valence states of Ni$^{2+}$ ($3d^8$) of La$_2$NiO$_4$ ($n$=1) and Ni$^{3+}$ ($3d^7$) of infinite layered LaNiO$_3$ ($n$=∞) [15], which behaves as $3d^{7.5}$ state. However, recent AP XPS results evidence no Ni$^{2+}$ and Ni$^{3+}$ charge disproportionation [21], instead supporting the admixtures of $d^8$ and $d^8\underline{L}$ ground state with $\underline{L}$ representing a ligand hole. At HP, BL-La327 undergoes structural phase transitions (SPT) from orthorhombic $Amam$ symmetry to $Fmmm$ symmetry (space group No. 69, $Z$=4) [15] for $P \geq 10.0$ GPa at room temperature (RT), eventually to $I4/mmm$ symmetry (space group No. 139, $Z$=2) for $P \geq 15.0$ GPa at RT or $P \geq 19.5$ GPa at 40 K [22]. Schematic crystal structures of $Fmmm$ symmetry and $I4/mmm$ symmetry are also shown in Fig. 1(a). All crystal structures were plotted by VESTA [23]. After SPT, the bond angle between Ni-O$_{iap}$-Ni changes to 180° to significantly enhance the interlayer coupling, and the in-plane Ni-O bonds become equivalent. Taken together with the shrinkage of bond length of Ni-O [15,22], the distortion of NiO$_6$ is gradually reduced for pressurized BL-La327. It also indicates the decline of CFS of Ni electronic states, which may have certain impacts on the SC like cuprates [14].

Soon after the discovery of pressure-induced high-$T_c$ SC in BL-La327 nickelates, many experimental and theoretical works have been reported to understand its electronic band structure [21,24-27], normal state properties [28,29], and SC [30-44].

Experimentally, ARPES measurements show that the $d_{x^2-y^2}$ and $d_{3z^2-r^2}$ bands dominate near $E_F$, and the former one cross the Fermi level forming electron-like and hole-like sheets [25,26]. However, $d_{3z^2-r^2}$ bonding band is immerged below $E_F$. These results are in agreement with theoretical calculations [30-32]. Due to the stronger electronic correlation, $d_{3z^2-r^2}$ electrons are likely localized while itinerant for $d_{x^2-y^2}$ electrons [38,45]. Optical conductivity study also shows the multiband nature of BL-La327 with dominated $d_{x^2-y^2}$ and $d_{3z^2-r^2}$ orbitals near $E_F$, and the $d_{3z^2-r^2}$ band sinks below Fermi level when temperature lowers below $T^* \cong 115$ K [27]. Study of X-ray absorption spectroscopy (XAS) and resonant inelastic X-ray scattering (RIXS) on BL-La327 single crystal has revealed that the $d_{x^2-y^2}$ and $d_{3z^2-r^2}$ orbital of Ni, and $2p$ orbitals of ligand oxygen dominate the low-energy physics with a small $\Delta_{CT}$ of 0~2 eV at AP [24]. Also, spin density wave (SDW) transition and magnetic excitation are demonstrated at AP [24,29], which exists a much stronger inter-layer effective magnetic superexchange interaction than those of intra-layer [24]. Moreover, experimental results based on multislice electron ptychography technique assisted by electron energy-loss spectroscopy (EELS) have directly demonstrated the oxygen vacancies mainly results from the $O_{iap}$ sites in BL-La327 with 2222-stacking order [46], which has strong CT character with ligand holes distributing at the inner apical and planar oxygen sites. Therefore, BL-La327 seems to possess a dominant ground state of $d^8$ and $d^8\underline{L}$ admixtures at AP [21,24,46]. In addition, the comparable $\Delta_{CT}$ of BL-La327 to cuprates also hints a favor factor for high-$T_c$ SC [12].

Current experimental results evidence the features of orbital dependent strong electronic correlations and a significant interlayer coupling due to $O_{iap}$ atoms in BL-La327 at AP [24-27]. Theoretically, it is pointed out that strong hybridization exists between $d_{3z^2-r^2}$ and $d_{x^2-y^2}$ orbitals [30,31,33,37,47], as induced by in-plane nearest-neighbor Ni sites through the Ni-O-Ni bond [33]. The bilayer structure is believed to be one crucial factor for SC in pressurized BL-La327 [30,33], which creates an interlayer superexchange AFM interaction of $d_{3z^2-r^2}$ electrons through the $O_{iap}$ $p_z$ orbital to induce local spin singlets with large pairing energy in strong-coupling limit [33]. While, phase coherence can be established through the increased orbital hybridization with the itinerant $d_{x^2-y^2}$ orbitals at HP. The superconducting pairing scenario due to stronger correlated $d_{3z^2-r^2}$ electrons is also reported in other studies [34,37,39].

On the other hand, other strong-coupling theory considers the onsite Hund's coupling of Ni with forming an interorbital spin-triplet state as well as the interlayer superexchange AFM interaction of $d_{3z^2-r^2}$ electrons in the limit $J_H \gg J_{\parallel,\perp}$ ($J_H$ is Hund's coupling, $J_{\parallel,\perp}$ represents intralayer and interlayer AFM spin-exchange) [38,42], which readily transmits to $d_{x^2-y^2}$ electrons. It is also shown that the HP BL-La327 is a multiorbital Hund metal with orbital-selective electronic correlation [45]. Benefitting from the Hund's coupling, the enhanced interlayer superexchange AFM interaction could provide the interlayer pairing and strongly enhances the pairing strength in $d_{x^2-y^2}$ orbital [35,38,42,44].

Another debate is whether the emerging $\gamma$ hole pocket from the bonding $d_{3z^2-r^2}$ band cross $E_F$ is at play in SC of BL-La327 at HP. Originally, the emergence of $\gamma$ hole pocket signals the $\sigma$-metallization due to its hybridization with the itinerant $d_{x^2-y^2}$ orbitals [15,30], which gives the high-$T_c$ SC in BL-La327. However, appearance of $\gamma$ hole pocket is argued to cause strong ferromagnetic (FM) fluctuations [40], as is destructive for singlet pairing. Recent Hall effect measurements show that the Hall coefficient $R_H$ undergoes a clear drop at about 10 GPa [17,48], signaling a notable increase of hole carrier density. This implies a change of electronic band structure topology. Concomitantly, SC appears as the anomaly sets in $R_H(P)$. It seems to be well consistent with the $\sigma$-metallization scenario [15,30], which naturally add more itinerant holes by hybridizing with in-plane $d_{x^2-y^2}$ bands.

However, current experimental studies of electronic band structure are largely limited to AP conditions, those of BL-La327 at HP remain absent and open for experimental studies. On one hand, the $\Delta E$ (primarily governed by apical O atoms) value between $d_{x^2-y^2}$ and $d_{3z^2-r^2}$ orbitals is one key parameter in determining both the hybridization and the Fermi surface topology in cuprates [14], the magnitude of which is essential for SC. On the other hand, it is known that the elongation of Ni-O$_{ap}$ bond is the most severe cause of the NiO$_6$ octahedrons distortion in BL-La327. By XRD measurements, it is demonstrated that pressure can greatly eliminate this distortion [15,22]. In this context, the evolution of $\Delta E$ under pressure and its connection to the distortion degree of NiO$_6$ octahedrons are crucial for understanding the pressure-induced SC in BL-La327, which motivates the present study. Herein, we report a systematic study of electronic band structures for both SL and BL RP-nickelates, which are probed by *in-situ* HP X-ray absorption near edge spectroscopy (XANES). Generally speaking, Ni $L$-edge directly probes the $3d$ orbital unoccupied states by dipole transitions of $2p_{1/2} \rightarrow 3d_{3/2}$ ($L_2$-edge) and/or $2p_{3/2} \rightarrow 3d_{5/2}$ ($L_3$-edge) while $1s \rightarrow 3p$ transitions for Ni $K$-edge, which indirectly provide the $3d$ orbital features through the pre-edge peak. The pre-edge peak is sensitive to the ligand coordination environments [49,50], thus the CFS. Meanwhile, the $\Delta E$ of main edge for $K$-edge XANES is directly related to the crystal field symmetry and bond length between transition metal and ligand [51,52]. Due to the severe attenuation by diamond anvil, the soft X-ray used for $L$ edge can hardly penetrate through the normal diamond anvils cell (DAC) that generates HP condition. Therefore, HP Ni $K$-edge XANES spectra at about 8333 eV were measured in this work.

Experimentally, the absorption coefficient $\mu(E)$ was obtained by two modes, i.e., energy-dispersive mode at ODE beamline [53], synchrotron SOLEIL, and traditional scanning energy mode using upstream ion chamber and downstream ion chamber at I18 beamline [54], Diamond Light Source Ltd. Simplified schematic setup is displayed in Fig. 1(b). We have collected the AP and HP XANES data for three different samples, denoted as SL-La214-ssr, BL-La327-ssr, and BL-La327-sol-gel, where ssr represents that the sample was synthesized by solid state reaction (ssr) method while sol-gel method for the third sample. As shown in Fig. S1 [55], the phase purity was characterized by Cu-$K_\alpha$ radiation, and Rietveld refinements results are given in table

S1 in supplementary materials [55]. Moreover, they display distinguishing magnetic and electrical transports properties at AP and HP [Fig. S2 and Fig. S3 [55]]. Both SL-La214-ssr and BL-La327-sol-gel samples show a strong and weak semiconducting behavior at AP and HP, and a subtle resistivity drop at about 80 K is observed in the later one at around 20.1 GPa. For BL-327-ssr, it is metallic down to 126.5 K at AP [16], below which it becomes weakly semiconducting. This behavior is agreement with that of BL-La327 with excess oxygen [20]. With applying pressure, metallization and strong resistivity drop at about 80 K are observed above 15 GPa. More details on the sample synthesis and HP characterizations are reported in ref. [16].

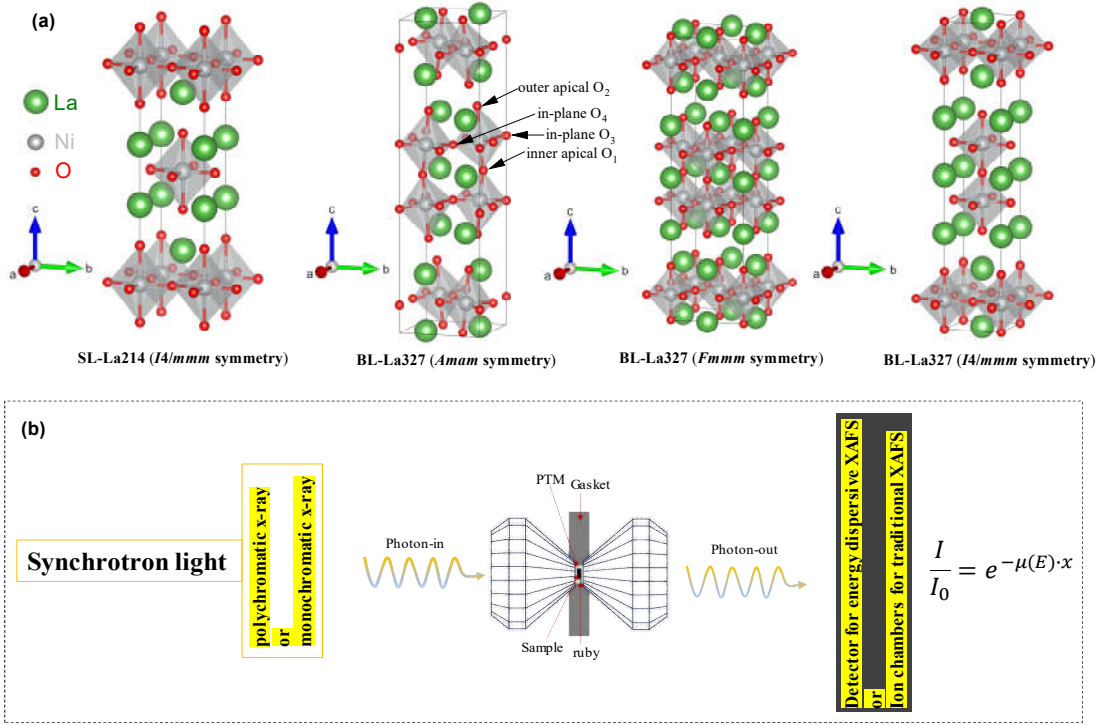

**FIG. 1. Crystal symmetries and experimental HP XANES setup.** (a) Crystal structures of SL-La214 at AP and BL-La327 at AP and HP. Four oxygen sites of BL-La327 at AP are indicated by arrows. (b) HP setup schematic for XAFS measurements using energy-dispersive mode or traditional energy scanning mode with ion chambers to obtain XANES data.

## II. RESULTS

### A. Electronic band structure and projected density of states

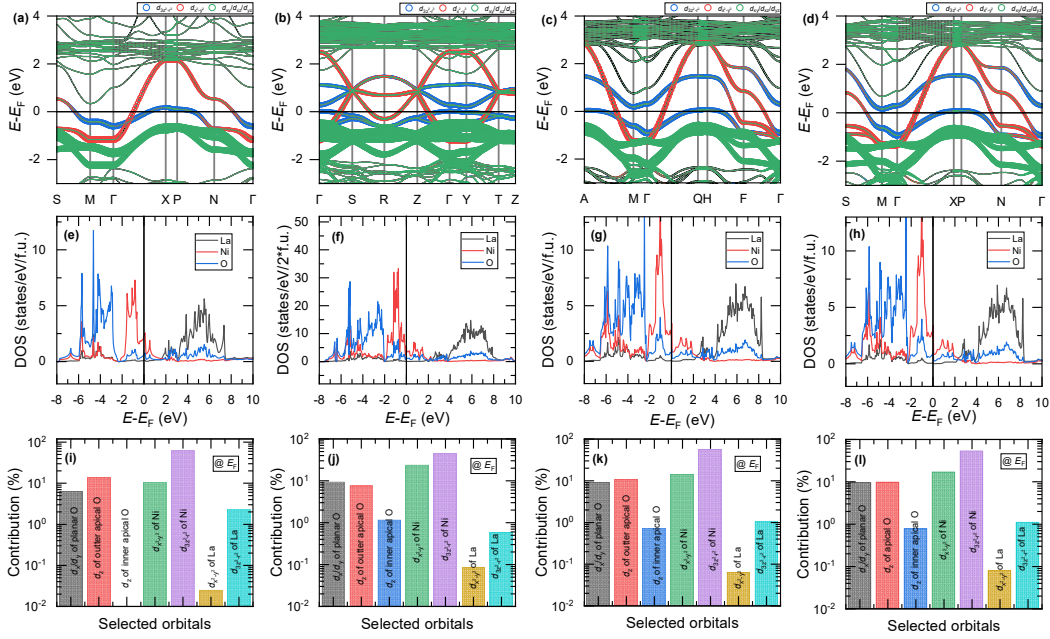

**FIG. 2. Theoretical electronic band structures and PDOS.** (a)-(d) are electronic band structures of SL-La214 ($I4/mmm$ phase at AP), BL-La327 ($Amam$ phase at AP), BL-La327 ($Fmmm$ phase at 29.5 GPa), and BL-La327 ($I4/mmm$ phase at 19.6 GPa). (e)-(f) are corresponding PDOS. (i)-(l) are corresponding DOS contribution to total for selected orbitals at $E_F$.

Before the experimental sections, we first present the results of theoretical electronic band structures for SL-La214 and BL-La327 nickelates, which are closely relevant to the experimental XANES features. From DFT-based calculations, we have obtained the electronic band structure for tetragonal La$_2$NiO$_4$ with $I4/mmm$ symmetry, orthorhombic La$_3$Ni$_2$O$_7$ with $Amam$ symmetry, orthorhombic La$_3$Ni$_2$O$_7$ with $Fmmm$ symmetry, and tetragonal La$_3$Ni$_2$O$_7$ with $I4/mmm$ symmetry [Figs. 2 (a)-(d)]. Schematics of primitive unit cell and first Brillouin zone can be found in Fig. S4 [55]. Calculations were performed without including effective Hubbard term $U_{eff} = U - J$ ($U$ is on-site Coulomb term and $J$ is site exchange term). For comparison, we also conducted the DFT + $U$ calculations with $U_{eff}$=4.0 eV, in which the $U$ term mainly affects the bandwidth of $d$-orbitals derived bands below $E_F$ but not observably above $E_F$. Additionally, $U$ term also renormalizes the projection weights and enlarges the energy splitting between $d_{3z^2-r^2}/d_{x^2-y^2}$ and $d_{xy}/d_{xz}/d_{yz}$ orbitals below $E_F$. Overall, it will not change the main features of the electronic band structures, as is consistent with other reports [17,25,26,46]. More details can be found in Fig. S7 [55].

Comparing to SL-La214, one distinguishing feature of BL-La327 is the emergence of new $d_{3z^2-r^2}$ dominant bands above $E_F$ between $d_{3z^2-r^2}$ bonding and $d_{x^2-y^2}$ derived bands. This is largely due to the existence of inner apical O atoms in BL-La327, which connect the two NiO$_6$ octahedron to enhance the Ni-O interlayer superexchange AFM interaction. The newly formed $d_{3z^2-r^2}$ band is antibonding one due to half filling feature with stronger electronic correlation strength than $d_{x^2-y^2}$ derived bands [15,25], which have a quarter electron filling. The second noticeable feature is the hybridization between $d_{3z^2-r^2}$ and $d_{x^2-y^2}$ derived bands, which manifests the mixed projection. The hybridization enables the $d_{3z^2-r^2}$ band contributed carriers to become more itinerant, which is proposed to be one key ingredient for phase coherence to induce SC [30,33]. All the $d_{xy}/d_{xz}/d_y$ derived bands are below $E_F$ since they are fully occupied by six electrons, which are not important for low energy physics. Additionally, the projections of O $p_x/p_y/p_z$ orbitals and La $3d$ orbitals to electronic band structure are presented in Fig. S5 and S6 [55], which show the orbital hybridizations among La, Ni, and O atoms at variable energy scale.

To investigate individual orbital contribution of La, Ni and O atoms, the total projected density of states (PDOS) are at first plotted to compare SL-La214 and BL-La327 [Figs. 2(e)-(h)]. From $E_F$ to -8 eV, the occupied states are dominated by the $3d$-orbitals of Ni atoms and $2p$-orbitals of O atoms. At $E_F$, the contributed DOS of selected orbitals to total are summarized as columns in Figs. 2(i)-(l), which are at play in transport and magnetism in reality. It turns out that the $d_{3z^2-r^2}$ and $d_{x^2-y^2}$ orbitals of Ni are primary contributors, and the $p_x/p_y/p_z$ orbitals of O are secondary, which supports the existence of ligand hole states in both nickelates. The contributions of La at $E_F$ is mainly from $5d_{3z^2-r^2}$ with less than 2.5% while that of $5d_{x^2-y^2}$ is negligible. Notably, there are two vital differences between SL-La214 and BL-La327 nickelates. One is that the DOS of Ni and O extends to about 1.0 eV above $E_F$ for SL-La214, while to 2.0 eV for BL-La327. The other one is that the strong hybridization between $2p$-orbitals of O and $5d$-orbitals of La and $3d$-orbitals of Ni occurs between about 2–3 eV for SL-La214, while 3–4 eV for BL-La327.

## B. XANES: experimental data and theoretical simulations

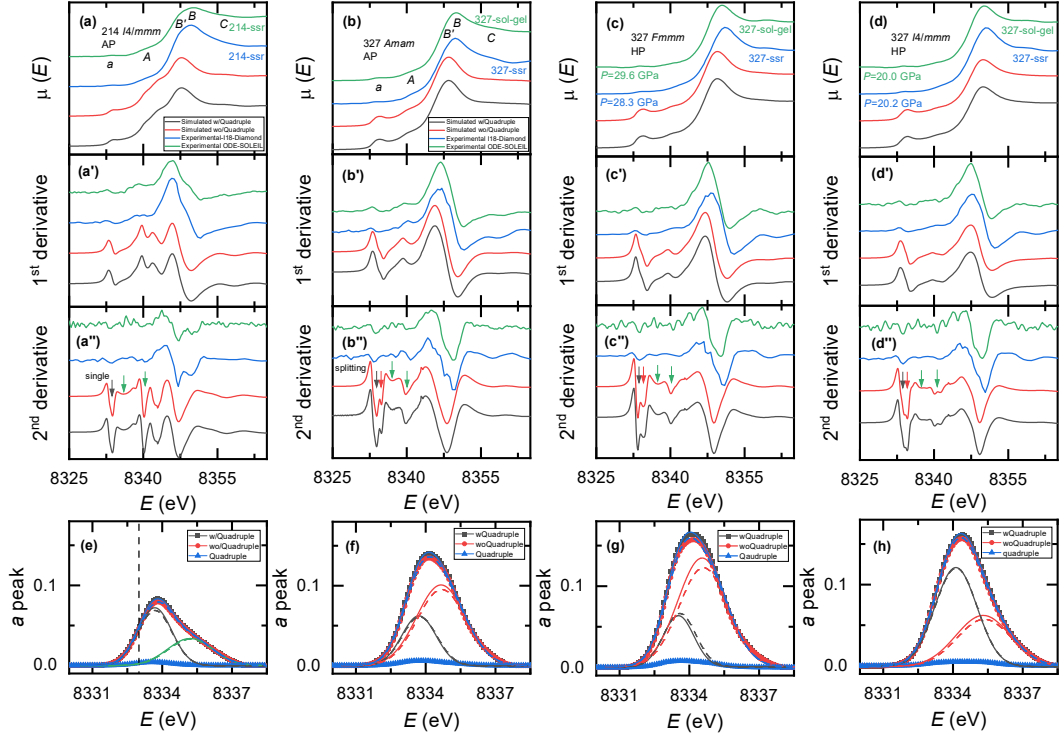

**FIG. 3. Theoretical and experimental comparisons of XANES data.** (a)-(d) are experimental and theoretical simulated XANES spectra. Alphabets of $a$, $A$, $B'$, $B$ and $C$ indicate the main features. (a')-(d') are corresponding first-order derivatives of XANES spectra. (a'')-(d'') are corresponding second-order derivatives. Arrows in (a'')-(d'') indicate the peak positions. (e)-(h) are quadruple contribution evaluated from fitting $a$ peak by Gaussian function based on the theoretical simulated XANES spectra with and without including quadruple channels.

XANES is a spectroscopic probe of density of empty states, providing electronic structure features above $E_\mathrm{F}$ of the target sample. For compounds containing transitional metal (*TM*), its pre-edge peak of $K$-edge XANES spectra mainly originates from electric dipole transitions ($1s \rightarrow np$) due to $p-d$ orbital hybridization and electric quadruple transitions ($1s \rightarrow nd$) with $n$ is shell number [49,50]. Using FDMNES code [56,57], we have conducted theoretical simulations of Ni $K$-edge XANES spectra of BL-La214 and BL-La327 with considering spin-orbital coupling (SOC). Electronic configurations of La, Ni, and O for XANES simulations are given in table S2 [55]. Together with the experimental XANES spectra, we plot them in Figs. 3(a)-(d) for comparisons. Correspondingly, the first and second derivatives are presented in Figs. 3(a')-(d') and Figs. 3(a'')-(d'') to determine edge position and energy splitting scale, respectively.

Experimentally, there are five peaks observed in both BL-La214 and BL-La327 at AP, labeled as $a$, $A$, $B'/B$ and $C$, which are consistent with previous reports [58,59]. Our simulations support that the $a$ peak mainly derives from dipole allowed transition $1s \rightarrow 4p_{hyb}^a$ due to the hybridization of Ni $3d/4p$ and O $2p$ orbitals as induced by the strong distortion of NiO$_6$ octahedron, and minor quadruple transition of $1s \rightarrow 3d$ associated with $d_{3z^2-r^2}$ and $d_{x^2-y^2}$ orbitals, as summarized in table S3 and table S4 [55]. As discussed in supplementary materials [55], group theory analysis gives no indications of $p-d$ orbital mixing in $D_{2h}$ and $D_{4h}$ symmetry for the non-orthogonalized terms unless considering the distortion of NiO$_6$ octahedron [49]. Group theory analysis is summarized in table S5 [55]. In fact, the elongation of the bond between Ni and apical O is the most severe distortion cause as demonstrated experimentally [15,22], resulting in two inequivalent O sites with inner apical O$_{iap}$ and outer apical O$_{oap}$. The bond distances between Ni-O$_{iap}$ and Ni-O$_{oap}$ are about 0.06 Å and 0.1 Å longer than those of in-plane Ni-O ones at AP. On the other hand, the oxygen vacancies that dominantly residing at O$_{iap}$ sites further break the local symmetry to promote the $a$ peak intensity. Through these two ingredients, the noncentrosymmetric NiO$_6$ octahedron is ought to significantly increase the intensity of $a$ peak besides the quadruple contributions [49], yet making $a$ peak intensity to be an indicator of $d$-orbital hole counts beyond distortions and vacancies factors.

With opening and closing the quadruple term, we are able to evaluate the quadruple contributions to $a$ peak by less than 5% as seen in Figs. 3(e)-(h), and table S3 [55]. It demonstrates a minor contribution of $1s \rightarrow 3d$ transitions, and the dominant dipole allowed $1s \rightarrow 4p$ transitions due to the enhanced $p$ orbital component of strong Ni-O orbital hybridizations, which is corroborated by theoretical band structure and PDOS [Figs. 2(a)-(h)]. More details on the atomic orbital projection of band structures and PDOS with $U$ and without $U$ term can be found in Figs. S5-S11 [55]. This enables us to probe the empty states of $d_{3z^2-r^2}$ and $d_{x^2-y^2}$ orbitals indirectly. One key finding in this work is that there is energy splitting of $a$ peak associated with Ni $3d$ orbitals for BL-La327 but not in SL-La214 below 8336 eV, as indicated by the arrows in Figs. 2(e)-(h). Relying on electronic band structure and PDOS calculations, we attribute it to the energy difference of the $d_{3z^2-r^2}$ and $d_{x^2-y^2}$ bandwidth at gravity center of BL-La327, which represents the CFS energy. For SL-La214, its $a$ peak is mainly caused by the $d_{x^2-y^2}$ band because of the negligible contribution of $d_{3z^2-r^2}$ to DOS above $E_F$. This is supported by the polarized XANES study of Sr doped SL-La214, which shows that $a$ peak is much more intense for $\sigma$-polarization ($\epsilon \perp c$) than $\pi$-polarization ($\epsilon \parallel c$) [59]. It directly illustrates that $a$ peak dominantly arises from the in-plane empty states, i.e., $d_{x^2-y^2}$ band.

In principle, when considering the core-hole effect in the final state [60,61], the $B'/B$ and $C$ peaks can be assigned by dipole allowed $1s \rightarrow 4p$ transitions associated with different ground states of Ni absorber. For $n$=1 with formal 2+ valence [61], the ground state of Ni can be written as $\alpha|3d^8\rangle + \beta|3d^9\underline{L}\rangle$ and $\alpha^2 + \beta^2 \leq 1$, and the former one is the leading state. The ligand hole states were verified by the O $K$-edge XANES spectra, which appears a strong pre-edge peak at around 528 eV [62]. The CT of ligand electrons of O to the localized $3d$ orbitals of Ni leaves a ligand hole state, which provides well screening of the core-hole but poorly screening for $3d^8$ state [60]. Because of the Coulomb interaction between the core-hole and the well localized $d$ electrons, the Coulomb energy gain pushes the $3d^8$ state to higher energy than the $3d^9\underline{L}$, which concomitantly changes their admixture in the final state [61]. Consequently, $B'/B$ and $C$ peaks have been ascribed to $1s \rightarrow 4p3d^9\underline{L}$ and $1s \rightarrow 4p3d^8$ transitions for SL-La214 [58].

In the case of $n=\infty$ with formal 3+ valence, due to the negative CT state [63,64], Ni therein is believed to have an admixture of $\alpha|3d^8\rangle + \beta|3d^8\underline{L}\rangle$ state with the later one as dominant [64-66]. Recently, XAS and RIXS experiments on BL-La327 single crystal suggests that it hosts a primary $3d^8$ ground state of Ni with non-negligible ligand holes for ambient $Amam$ phase [24], i.e., $\alpha|3d^8\rangle + \beta|3d^8\underline{L}\rangle + \gamma|3d^7\rangle$. In addition, O $K$-edge EELS studies of BL-La327 single crystal also prove the strong $p-d$ hybridization and strong CT nature of Ni-O [46]. The ligand holes mainly reside at the inner apical O and in-plane O sites [46], indicating both $d_{3z^2-r^2}$ and $d_{x^2-y^2}$ orbitals are evolved in low energy physics. In our theoretical simulations of Ni $K$-edge XANES spectra, we have considered a simplest situation with setting the initial electronic states of Ni to be $3d^84s^2$ while $3d^84s^24p^1$ for the final state. We find that the simulated XANES spectra unexpectedly match the experimental ones quite well, and all the $a$, $A$, $B'/B$ and $C$ are observed [Figs. 2(a)-(d)].

According to the electronic band structure and PDOS results, we suggest that the most important ingredient to enhance the weak features of $a$, $A$, and $C$ are the strong $p-d$ hybridization of among La-Ni-O atoms, which open new dipole-allowed transition channels yet carrying the $d$ orbitals components of Ni and La, especially for the former two. The enhancement should be insignificant for $C$ peak due to the negligible $p-d$ hybridizations among La-Ni-O atoms, as seen in Figs. S8-S10 [55]. Noteworthy that, the intensity is still much weaker than the white line at $B$ since the wavefunction overlapping integral is small at energy scale of $a$, $A$ and $C$ peaks due to the minor DOS of Ni $p$ orbital. For this reason, the $A$ peak can be mainly attributed to dipole allowed $1s \to 4p$ transitions enhanced from the $p-d$ hybridizations of outer apical O $2p$ and La $5d$ orbitals [Figs. 2(e)-(h)]. The interpretation is consistent with the results from the outer apical O $K$-edge EELS spectra simulated by DFT calculations, in which the peak at 531.1-532.8 eV is ascribed to the hybridization of outer apical O $p$ orbitals and La $5d$ orbitals [46]. Benefit from the stronger lifting of the O $2p$ and La $5d$ orbitals derived bands, their contribution to $a$ peak becomes negligible in BL-La327 compared to that of SL-La214.

To explain the $B'/B$ and $C$ peaks, it is necessary to consider the multiplet of Ni in ground states. As seen in Figs. S8-S9 [55], there is clear splitting of the DOS for degenerated $4p_x/4p_y$ and $4p_z$ orbitals at 10–20 eV for both SL-La214 and BL-La327. It naturally explains the main edge splitting as assigned as $B'/B$ peaks, in agreement with the previous reports on the powder samples [58,59,61]. Similarly, the weak $C$ peak at 20–30 eV is caused by the $4p_x/4p_y$ and $4p_z$ orbitals without apparent splitting in DOS. We leave the discussion on the intrinsic splitting of the main edge in section D. Nevertheless, we note that the strong polarization-dependent of both main edge and weak $C$ peak observed in SL-La214 single crystal suggests other origins besides DOS splitting in the $RP$ nickelates. On plausible scenario is the core-hole screening effect associated with the mutiplet of Ni $3d$ ground electronic configurations [60,61]. Similar to SL-La214 [58], we propose that the $B'/B$ and $C$ peaks dominantly result from the $1s \to 4p3d^8\underline{L}$ and $1s \to 4p3d^8$ transitions, as summarized in table S4 [55].

## C. Pressure dependent Fermi energy $E_\text{F}$ and $d$-orbital energy splitting

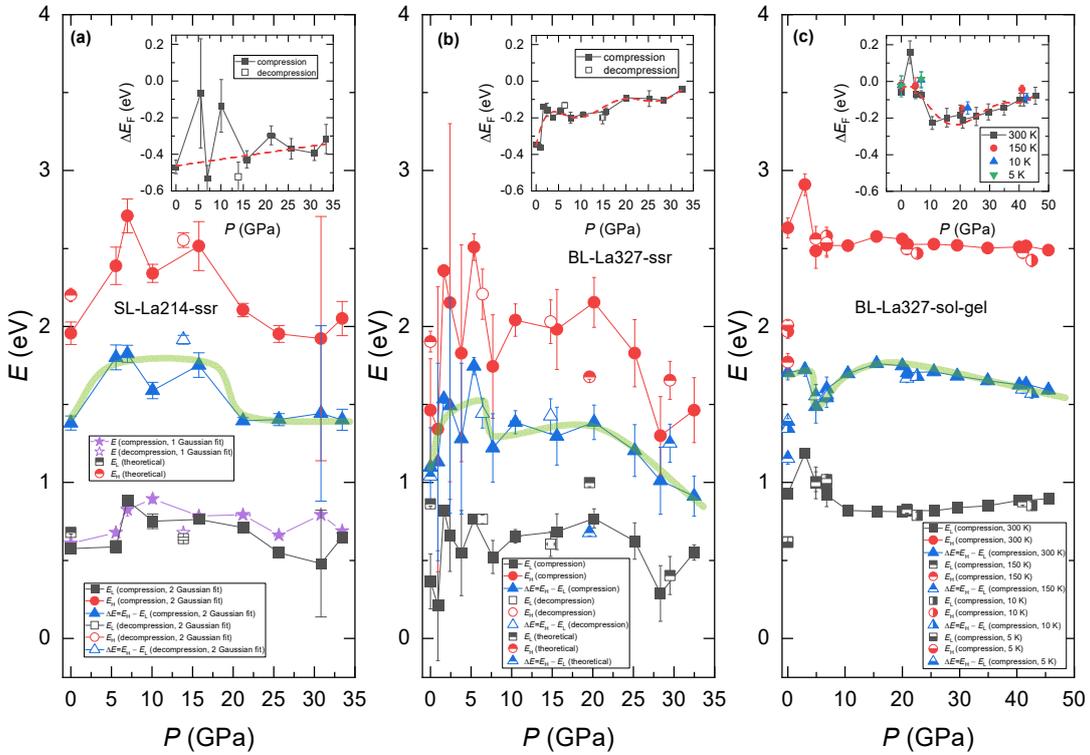

**FIG. 4. Experimental evidences of the $d$-orbital energy splitting.** Energy splitting of $a$ peak as a function of pressure for (a) SL-La214-ssr sample, (b) BL-La327-ssr sample, and (c) BL-La327-

sol-gel sample. Vertical axis of main panels and insets are pressure-dependent relative energy ($E$) and Fermi energy $\Delta E_F$ to that of Ni foil at $E_0$ =8333 eV. Solid symbols are for compression, empty symbols for decompression, and half-down empty symbols for theoretical simulations, and others in (c) for low temperature. Dashed red lines in insets and solid green lines are guide to eyes.

Generally, it is not an easy task to determine the valence state for highly covalent compounds [60], like *RP*-nickelates with multiplet ground states [21,24,58,61,62,64-66]. Technically, this is largely because of the quite weak nature of pure electric quadruple transitions and their mixing with $p-d$ hybridization enhanced dipole transitions at *TM K*-edge XANES. Also, besides oxidation state, the spin state, geometry and others could also vary the energy position, splitting, and intensity distribution of the pre-edge peak [49]. Instead, one properly uses the multiplet of the ground states configurations of *TM*s to describe the low energy physics in strongly correlated systems. This will be deeply discussed in the last section for SL-La214 and BL-La327 nickelates.

Since $a$ peak is closely related to the empty states above $E_F$, we have analyzed the pressure dependence of $E_F$ as determined by approximating the first derivative peak using Voigt function. The evolution of $E_F$ could be a good approximate for evaluating the relative valence state change of Ni in SL-La214 and BL-La327 under pressure. The results are summarized in the insets of Figs. 3(a)-(c). We find that there is only a small increase of $E_F$ for SL-La214-ssr, and its relative change at maximum pressure (*P*=33.4 GPa) to ambient $\Delta E_F$ reaches about 0.15 eV. For BL-La327-ssr sample, $\Delta E_F$ reaches to 0.30 eV at 32.5 GPa, which is twofold of SL-214-ssr. However, $\Delta E_F$ reaches to -0.021 eV for BL-327-sol-gel sample. These experimental facts clearly suggest the distinguishing electronic band structure at around $E_F$.

The $T_c$ is suggested to positively correlate with the energy splitting $\Delta E$ of the $d_{3z^2-r^2}$ and $d_{x^2-y^2}$ bands in SL cuprates [14], which is primarily governed by the distance between Cu and apical O atom. To investigate the pressure dependence of $\Delta E$ associated with antibonding $d_{3z^2-r^2}$ and $d_{x^2-y^2}$ bands, we have fitted the $a$ peak using two Gaussian functions. The $a$ peak is obtained by subtracting the background as is approximated by cubic spine line fitting [58], as displayed in Fig. S12 for theoretical spectra [55] and Fig. 8, Fig. 9, and Fig. 10 for experimental spectra in Appendix. We present the results of the three samples in Figs. 3(a)-(c). As discussed previously, the peak at lower energy ($E_L$) is mainly contributed by in-plane $d_{x^2-y^2}$ band for SL-La214-ssr sample, which has a weak pressure dependence. It first shows an increase till around 10 GPa, at which it begins to decline towards $E_F$. The La-O derived hybridized band at higher energy ($E_H$) has similar trend.

Figures 3 (b) and (c) show the $E_L$ ($d^*_{3z^2-r^2}$ and $d_{x^2-y^2}$ overlapped band) and $E_H$ ($d_{x^2-y^2}$ dominated band) evolutes with applying pressure as well as their difference as an estimate of the CFS energy $\Delta E_{cf} = E_H - E_L$ for two BL-La327 samples. At AP, $\Delta E_{cf}$ is about 1.10±0.25 eV for BL-327-ssr, in good agreement with theoretical simulated 1.04±0.04 eV for *Amam* phase. With applying pressure, both $E_L$ and $E_H$ tends to increase till about 5.3 GPa, above which it starts to decrease. Between 7.7 GPa and 20.2 GPa, they gradually grow before suffering a fast decrease. Simultaneously, the $\Delta E_{cf}$ behaves as the same trend. We also simulated the XANES of *Fmmm* phase at 29.5 GPa [15], and obtain $\Delta E_{cf}$=1.25±0.12 eV as close to the experimental value. However, a smaller $\Delta E_{cf}$=0.68±0.02 eV is obtained for *I4/mmm* phase at 19.6 GPa using the structure parameters refined from the HP XRD patterns at 40 K [22]. Notably, the shrinking of the CFS energy is well consistent with the reduction of the distortion of NiO$_6$ octahedra above 20 GPa [22]. If the scenario of $T_c$ positively correlating with $\Delta E$ also works in *RP*-nickelates, our results of $\Delta E_{cf}(P)$ would imply the synergetic role of intralayer and interlayer interaction of Ni and O atoms in superconducting pairing in BL-La327.

As for the BL-La327-sol-gel sample, $\Delta E_{cf}$ is about 1.70±0.05 eV at AP, much higher than BL-La327-ssr. This implies a severer NiO$_6$ distortion in BL-327-sol-gel than BL-327-ssr. Applying pressure above 3.0 GPa occurs a sharp drop of $\Delta E_{cf}$. Compared to BL-327-ssr, a more apparent increased of $\Delta E_{cf}$ is observed at 4.9-15.5

GPa before tending to decrease till 45.5 GPa. The data at 5 K, 10 K, and 150 K are complied with 300 K at HP while the low temperature $E_H$, $E_L$ and $\Delta E_{cf}$ seem to shift towards $E_F$ at AP, which may be related to the density wave (DW) transition at 150 K. More detailed study on this will be reported elsewhere. Above 20 GPa, we find one remarkably different feature of $d^*_{3z^2-r^2}$ and $d_{x^2-y^2}$ overlapped band between BL-La327-ssr and BL-La327-sol-gel. In other words, the $E_L$ starts to decrease quickly for the former one while gradually increases for the later one for $P \geq 20$ GPa. Overall, these distinguishing responses of electronic structure to pressure among the three samples have significant implications on understanding the pressure-tuned transport properties under nonhydrostatic compression [15,16].

## D. Pressure dependent main edge shift and *p*-orbital energy splitting

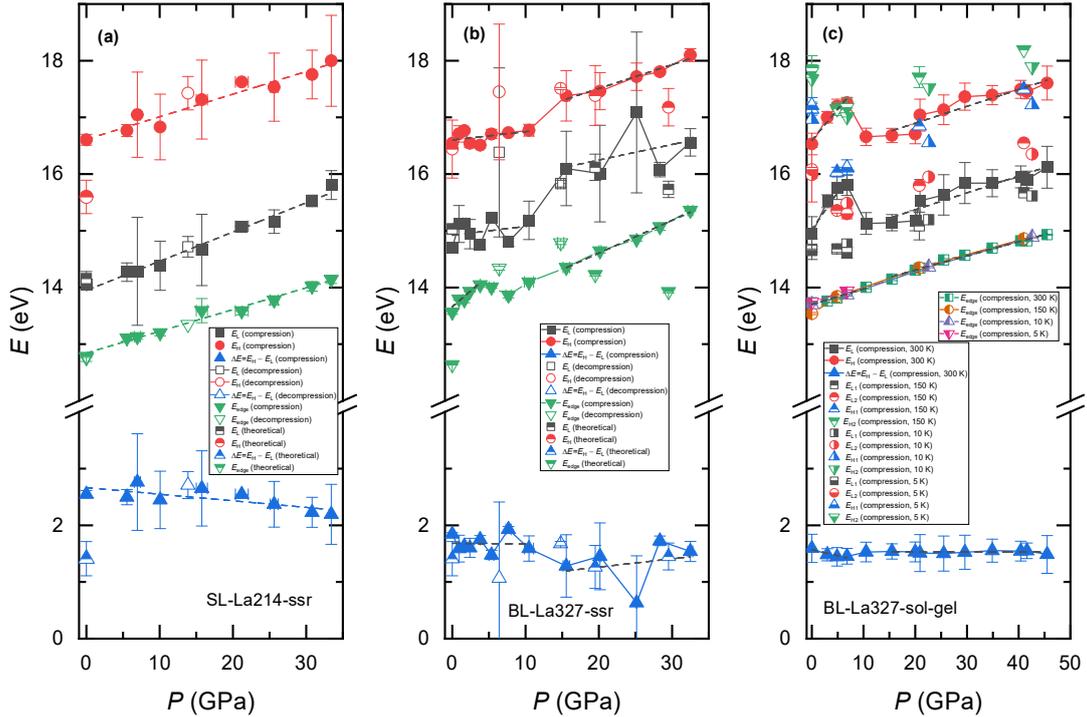

**FIG. 5. Experimental evidences of the *p*-orbital energy splitting.** Main edge energy $E_{edge}$ (green), $E_L$ (black), $E_H$ (red), and energy splitting $\Delta E$ (blue) of $B'/B$ peak as a function of pressure for (a) SL-La214-ssr sample, (b) BL-La327-ssr sample, and (c) BL-La327-sol-gel sample. All the energy values are reduced to that of Ni foil at $E_0 = 8333$ eV. Solid symbols are for compression, empty symbols for decompression, and half-down empty symbols for theoretical

simulations, and others in (c) for low temperature. Dashed lines in are linear fittings.

Now, we come to discuss the results of the pressure dependent main edge and white line splitting caused by the crystal field. As discussed in previous section, we recall here that the Ni $K$-edge of SL-La214 and BL-La327 are suggested to arise from the $1s \rightarrow 4p3d^9\underline{L}$ and $1s \rightarrow 4p3d^8\underline{L}$ transitions, respectively. As a result, both the crystal field and core-hole effect screened by ligand hole contribute to the actual energy splitting of the main edge. Theoretically, since the absorption coefficient $\mu(E)$ is proportional the product of the squared modulus of the transition matrix element and DOS [67], $\mu \propto \left|\langle \psi_f | \hat{\epsilon} \cdot \vec{r} e^{i\vec{k}\cdot\vec{r}} | \psi_i \rangle\right|^2 \rho(E_f - E_i - \hbar\omega)$, where $\psi_i$ and $\psi_f$ are initial and final electronic states at energy $E_i$ and $E_f$, while $\hat{\epsilon}$, $\vec{k}$, and $\hbar\omega$ are X-ray's electric polarization vector, momentum and energy, $\vec{r}$ is position operator, and $\rho$ is empty DOS. In the dipole approximation, it is written as $\mu \propto \left|\langle \psi_f | \hat{\epsilon} \cdot \vec{r} | \psi_i \rangle\right|^2 \rho$. As displayed in Figs. S8 and S9, the Ni $K$-edge splitting is readily resolved from the calculated PDOS of $p_z$ and degenerated $p_x/p_y$ orbitals with a splitting energy scale of around 2 eV at gravity center for SL-La214, and BL-La327 (i.e., ~2.32 eV, ~2.02 eV, 2.06 eV, ~2.27 eV, for SL-214, and BL-327's three phases). Within energy resolution (~1.17 eV at Ni $K$-edge) of X-ray light in this work, this enables us to well resolve the main edge splitting even in our powder samples. As shown in Fig. 8, Fig. 9, and Fig. 10 in appendix, the main edge ($E_{edge}$) and peak splitting energy ($\Delta E = E_H - E_L$) do manifest experimentally by the doublet peaks, as can be distinguished by the 1st/2nd derivatives of $\mu(E)$ [Fig. 2].

To determine $\Delta E$, we have fitted the doublet peaks in the 2nd derivatives by two Voigt functions to obtain the peak positions of $B'$ and $B$ peaks, denoted as $E_L$ and $E_H$, respectively. Meanwhile, the edge energy ($E_{edge}$) mostly representing that of the $B'$ peak is also extracted. According to Natoli's rule [52], when the XANES peaks at the scale of 13.6 eV$\leq E_p - E_0 \leq$40-163.2 eV, the peak energy position ($E_p$) would largely correlate with the inverse squared distances ($R^{-2}$) between absorber and its neighbor atoms, i.e., $(E_p - E_0)R^{-2}$=constant. In the case of present study, this implies that $\Delta E \propto R_c^{-2} - R_{ab}^{-2}$ among the Ni-O bonds, where $R_c$ ($R_{ab}$) corresponds to the bond distances between Ni and apical (in plane) O atoms, $E_0$ is constant potential. It is worth noting that, the peak energy $E_L$ and $E_H$ of $a$ peak are fairly insensitive to the bond distances of Ni and O [52], thus mostly reflecting the electronic states of Ni absorber. Natoli's rule has been verified to satisfy in cuprates and nickelates [58,61,68].

Experimental data are summarized in Figs. 4(a)-(c). With applying pressure, all the $E_L$, $E_H$ and $E_{edge}$ values tend to increase linearly for SL-La214-ssr. Concomitantly, $\Delta E$ is decreasing up to $P$=33.4 GPa due to the faster increasing rate of $E_L$. This indicates the effective compression on the distance between Ni and apical O is stronger than that of in plane Ni-O distance. However, the monotonic decrease of $\Delta E$ suggests no pressure-induced SPT up to 33.4 GPa. We note that Sr doping in SL-La214 (La$_{2-x}$Sr$_x$NiO$_4$, $0 \leq x \leq 1.6$) also induces much more significant shift of $E_L$ than $E_H$ towards to high energy [58]. The Sr dopants therein not only induce hole carriers above $x >0.6$ but also trigger a monotonically decline of semiconductor-metal transition temperature at doping level $0 \leq x \leq 1.2$ [2]. HP electrical transport measurements up to 33.5 GPa show that the resistivity of SL-La214-ssr remains highly semiconducting besides a small increase of conductivity [Fig. S2(a) [55]].

For the two BL-La327 samples, all the $E_L$, $E_H$ and $E_{edge}$ values display a nonmonotonic behavior, which manifests a contrasting response to pressure at $P$=5-15 GPa. In detail, both $E_L$, $E_H$ increase fast after applying pressure at $P <2.0$ GPa for BL-La327-ssr and $P < 4.9$ GPa for BL-La327-sol-gel, signaling a dramatic compression of lattice parameters. This is demonstrated by plotting the inverse squared Ni-O bond distances as well as the difference $\bar{R}_{ab}^{-2} - \bar{R}_c^{-2}$ as a function of pressure using the averaged lattice parameters [Fig. S13]. With further increasing pressure, $E_L$ and $E_H$ quickly decrease before showing a moderate pressure dependence up to 15.2 GPa for BL-327-ssr and 20.0 GPa for BL-327-sol-gel. Above 15.2 GPa and 20.0 GPa, $E_L$ and $E_H$ emerge a much more sudden increase for BL-327-ssr than that of BL-327-sol-gel. These strongly indicate the Ni-O bond length changes associated with the pressure-induced SPT with the sequence of $Amam \rightarrow Fmmm \rightarrow I4/mmm$ from the HP XRD studies [15,22]. The evolution of $\Delta E$ for BL-327-ssr is much more comply with that of $\bar{R}_c^{-2} - \bar{R}_{ab}^{-2}$ extract from previous report [15]. The drop of $\Delta E$ above 10.5

GPa clearly supports the reduction of the distortion of NiO₆ octahedron through faster compression along $c$-axis than in-plane.

However, the situation seems to be opposite for the BL-327-sol-gel sample. The $\Delta E$ instead shows an increase trending above 4.9 GPa, and it keeps a slight decline trending between 10.5 GP and 45.5 GPa. Therefore, it corresponds to a balanced compression on NiO₆ octahedron with almost parallel increase of the $E_L$ and $E_H$ as a function of pressure. Notably, both $E_L$ and $E_H$ split to two values at low temperatures, respectively. This implies a further symmetry lowering of NiO₆ octahedron, which may be related to the DW transition at about 150 K [29,69]. However, the $E_{edge}$ shows a negligible change of magnitude at HP and low temperature despite those at AP, which display an apparent temperature-dependence.

**E. Evidence of pressure-induced hole doping**

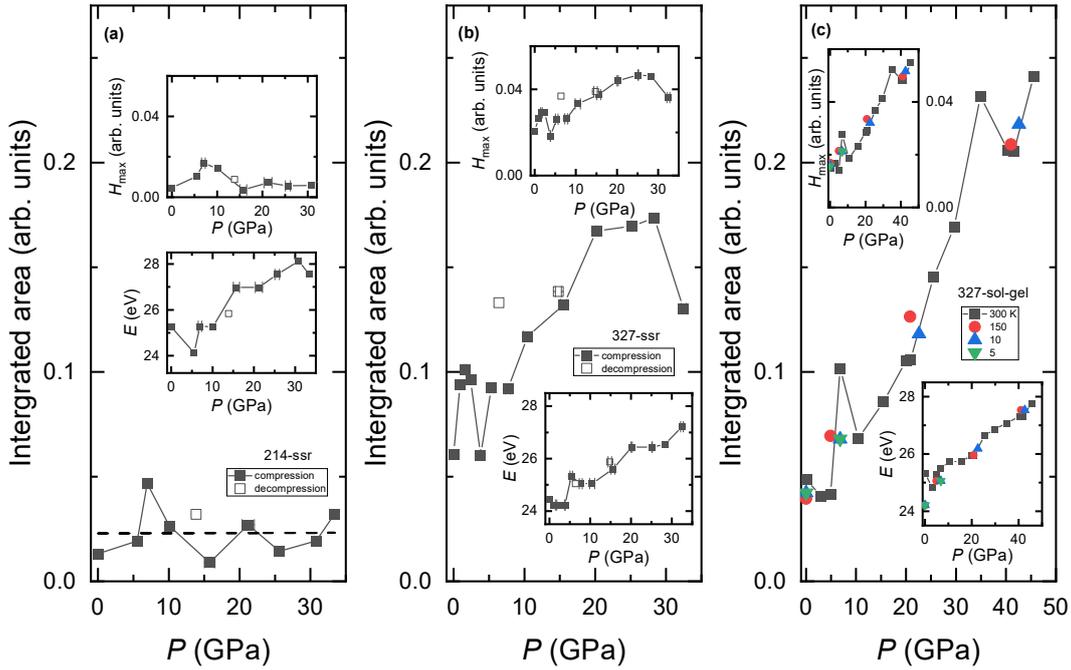

**FIG. 6. Experimental evidences of pressure-induced hole doping.** Integrated peak area of $C$ peak as a function of pressure for (a) SL-La214-ssr sample, (b) BL-La327-ssr sample, and (c) BL-La327-sol-gel sample. Insets are pressure dependence of maximum height and peak energy. All the energy values are reduced to that of Ni foil at $E_0$ =8333 eV. Solid symbols are for compression, empty symbols for decompression.

Previous report has shown that Sr-doping induces hole carriers in SL-La214 nickelate and Hall coefficient $R_H$ changes sign from negative to positive above doping content *x*=0.6 [2]. Concomitantly, the hole doping manifests in XANES as the increase of integrated peak area of $a$ peak and $C$ peak in Sr-doped SL-La214 [59]. We first analyze the $C$ peak and plot its integrated area, maximum height ($H_{max}$) and peak energy as a function of pressure in Figs. 6(a)-(c). The integrated area shows a weak pressure dependence with a slight increase up to 33.4 GPa for SL-214-ssr [Fig. 6(a)]. Its $H_{max}$ behaves as similarly. These indicate a rather inefficient hole doping effect by applying pressure below 33.4 GPa compared to chemical doping [2,59], which can strongly diminish the distortion of NiO$_6$ octahedra.

The semiconductor-metal transition disappears above Sr doping content around *x*=1.2 for SL-La214 [2], where the ratio of $R_c/R_{ab}$ declines to a minimum value. It is reasonable from the fact that $\Delta E$ of main edge has much weaker pressure dependence than that of Sr-doping [58], implying a pressure induced negligible reduction of NiO$_6$ octahedra distortion. The small changes of integrated area of $a$ peak [Fig. S13(a) [55]] and FWHM [Fig. 11(a)] further evidence the subtle hole doping effect via applying pressure up to 33.4 GPa [55]. Interestingly, the distance between apical O atom and La atom increases up to *x*=0.6, above which it becomes saturated. This indicates that the displacement of apical O atom controls the transition from localized electrons to itinerant electrons character at semiconductor-metal transition temperature [2]. Also, the Sr doping induces in-plane holes occupying $d_{x^2-y^2}$ orbital [58,59].

For BL-La327-ssr and BL-La327-sol-gel samples, the integrated area of $C$ peak shows an obvious increase trend for $P$ >7.7 GPa and $P$ >4.9 GPa, designated as a pressure-induced anomaly in electronic band structure. This trending also holds for $H_{max}$. The anomaly likely stems from the pressure-induced SPT from $Amam \rightarrow Fmmm$ or $I4/mmm$ as observed by HP XRD experiments [15,22]. Crossing the SPT, the NiO$_6$ octahedra distortion will be greatly eliminated through moving outer apical O atoms towards to Ni atoms and equalizing bond length between the in-plane O atoms and Ni atoms [15,22]. Similar to Sr doped SL-La214 [2], in the case of BL-La327, the pressure induced displacement of inner apical O atoms probably results in transition of localized holes to itinerant holes character, triggering the semiconductor-metal transition at HP. This can be realized by the enhanced hybridization between localized $d_{3z^2-r^2}$ orbital and itinerant $d_{x^2-y^2}$ orbital through interlayer interaction. Moreover, the sensitivity of $C$ peak position to lattice contraction is well evidenced by its nonmonotonic shift to higher energy, which relates to the distortion degree of NiO$_6$ octahedra [insets of Figs. 6(b) and (c)].

On the other hand, the pressure-induced hole doping causes the lowering of $E_F$, which may lead to the $\sigma$ bonding band of $d_{3z^2-r^2}$ cross $E_F$. This will contribute the hole counts and increase the hole carrier density, giving rise to the increased integrated area of both $a$ peak and $C$ peak. The integrated area of $a$ peak indeed shows an increase above 7.7 GPa, forming a dome-like shape up to 32.5 GPa for BL-La327-ssr and up to 45.5 GPa for BL-La327-sol-gel [Figs. S13 (b) and (c)]. Recently, the Hall effect measurements show a sudden increase of hole carrier density above about 10 GPa [17,48]. As will be presented in discussion section, we show that the conductivity positively correlates to the integrated area of $C$ peak [Fig. S14 [55]]. Taken together, we conclude that pressure can significantly induce hole doping in BL-La327 nickelates *via* evoking a SPT to diminish NiO$_6$ octahedra distortion.

Since the $C$ peak relates to transitions of $1s \rightarrow 4p3d^8$ and $1s \rightarrow 4p3d^7$ with the former one as the dominated term, the change of $C$ peak magnitude in height and area also hints its connection to the spin state of the $3d^8$ and $3d^7$ configuration [49]. It has been observed by HP resonant inelastic x-ray scattering (RIXS) and x-ray emission spectroscopy (XES) that there occurs transition from high spin to low spin of $3d^8$ configuration in Sr doped SL-La214 [70]. It manifests that the $H_{max}$ and integrated area of both $a$ peak and $C$ peak in low spin state at 57 GPa is much higher than that of high spin state at AP [70]. Through magnetic measurements [71], Kato et. al. suggest that there appears a two-dimensional AFM correlation between formal Ni$^{3+}$ ions with $S=$ 1/2 low-spin state when Sr doping level exceeds $x$=1.0 for La$_{2-x}$Sr$_x$NiO$_4$.

The very similar response to pressure of both $a$ peak and $C$ peak in BL-La327

direct us to speculate that there could exist a pressure-induced transition from high spin to low spin of $3d^8$ configuration or growing proportion of $3d^8\underline{L}$ and/or $3d^7$ states in BL-La327. The low spin state is consistent with the theoretical proposal on the electronic configuration of HP *Fmmm* phase [15], i.e., half filling on the $d_{3z^2-r^2}$ orbital while quarter filling on the $d_{x^2-y^2}$ orbital. These two configurations could mimic to low spin $3d^8\underline{L}$ or $3d^7$ states. Recent theory proposes that the ground state of BL-La327 exists an electronic correlation-driven spin crossover from a low-spin $S$=1/2 state to a high-spin $S$=3/2 state [36]. However, Jiang et. al. suggest that BL-La327 has a strong CT nature and its local bilayer Ni-O-Ni component is in an effective $Ni^{2+}$ $3d^8$ state with itinerant ligand holes primarily residing on in-plane O atoms at AP [43]. Applying pressure would greatly enhance the interlayer hopping to produce an exceptionally large interlayer superexchange interaction, which induces the fractional spin transition from effective $S$=1 state to $S$=1/2 state [43]. Without hole doping, ground state of $S$=1 Mott insulator is also proposed for BL-La327 [35]. Meanwhile, BL-La327 at HP is proposed to be a multiorbital Hund correlated metal with 33.28% high spin $S$=1 state with orbital occupancy $N$=2, 12.36% low spin $S$=1/2 with $N$=1 and 23.62% low spin $S$=0 state with $N$=2 based on the local multiplets at 290 K [45]. Theoretical calculations by Shilenko and Leonov show that HP *Fmmm* BL-La327 is close to a negative CT regime and its electronic configurations are estimated to possess 65% high-spin and 35% low-spin state [28]. These results strengthen the possibility of pressure-induced spin crossover transition in BL-La327, and quest its exact ground states at AP and HP. Our XANES results are mostly relevant to $3d^8$ with high spin $S$=1 and $3d^8\underline{L}$ with low spin $S$=1/2 scenario, suggesting the $C$ peak as an indicator of examining this transition hypothesis. Further HP RIXS at Ni K-edge and XES measurements are called for clarifying the occupied states and spin states of the BL-La327 nickelate.

## III. DISCUSSIONS

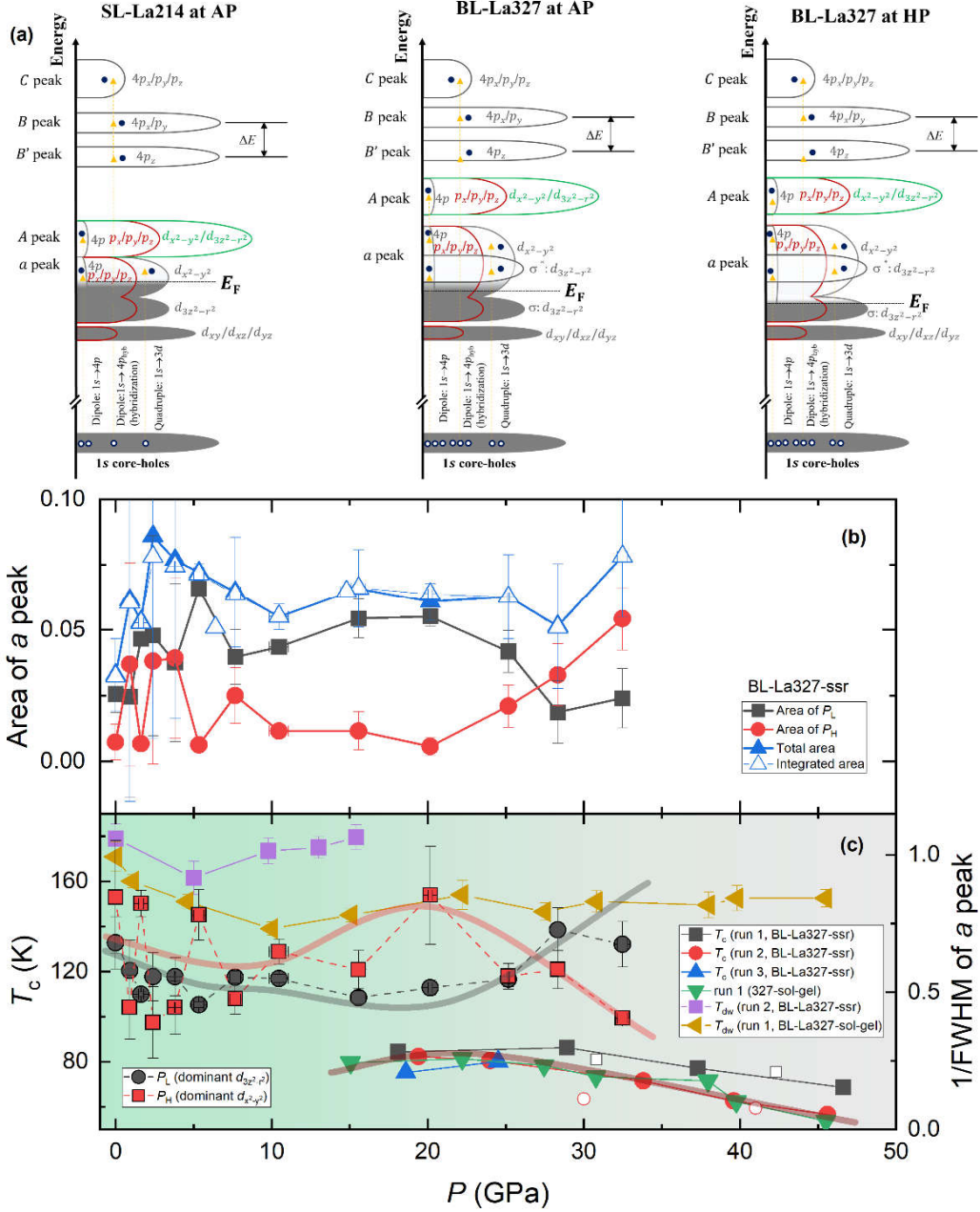

**FIG. 7. Phase diagram with schematic electronic transitions and $a$ peak area as a function of pressure.** (a) Schematics of electronic transitions of XANES process for SL-La214 and BL-La327. Green: La, Red: O, and Gray: Ni. (b) Pressure dependence of $a$ peak area for BL-327-ssr sample. (c) Pressure dependence of density wave-like transition temperature ($T_{dw}$) and superconducting transition temperature ($T_c$). Data of BL-La327-ssr are from ref. [16]. Open symbols are data for decompression. Pressure dependence of $1/\text{FWHM}$ for dominant $d_{3z^2-r^2}$ and $d_{x^2-y^2}$ orbitals is shown as the right axis. Solids lines are guide to eyes.

We first discuss the electrical transport properties of the three samples at AP and HP.

Temperature dependent resistivity $\rho(T)$ are plotted in Fig. S1 for AP and Fig. S2 for HP. At AP, the SL-La214-ssr shows a strong semiconducting behavior [Fig. S1(d)]. The semiconducting behavior remains up to 33.5 GPa with only a slight growth of conductivity at HP [Fig. S2(a)]. For BL-La327-ssr [Fig. S1(e) and Fig. S2(b) [55]], it shows metallic down to 126.5 K [16], below which it becomes weakly semiconducting. Applying pressure first causes the weak semiconducting behavior up to 15.4 GPa, which disappears at $P \geq 19.5$ GPa. It signals the occurrence of semiconductor-metal transition for pressurized BL-La327-ssr. Concomitantly, clear resistivity drop is also observed at about 80 K, which is associated with superconducting transition [16]. In contrast, BL-La327-sol-gel sample not only shows semiconducting behavior at AP, but also keeps matrix semiconducting behavior up to 45.5 GPa with only quite weak resistivity drop at about 80 K and 22.2 GPa [Fig. S1(f) and Fig. S2(c) [55]]. This implies much lower fraction of superconducting phase than that of BL-La327-ssr. In addition, there are two resistivity anomalies as may relate to the DW transition at high temperature and AP for BL-La327 [29,72], denoted as $T'_{dw}$=116.4 K and $T_{dw}$=179.0 K. Herein, the DW temperature is defined as the position of extrema of the first derivative of $\rho(T)$. With applying pressure, the $T'_{dw}$ is not extracted since it is smeared out. In the following we will focus on BL-La327-ssr due to its analogy to the reported HP electrical transport properties of superconducting single or polycrystalline samples [15,17,18].

To correlate the XANES feature with conductivity, we further compare with the pressure dependent integrated area of $C$ peak and conductivity. It is revealed that the conductivity has a positive correlation with the integrated area of $C$ peak [Fig. S13 [55]]. The positive correlation also exists in Sr-doped SL-La214 nickelates [2,58]. Moreover, it is known that there occurs metal-insulator transition (MIT) for the infinite layered (IL) IL-$R$113 ($R$=Pr, Nd, Sm, and Eu) except the metallic IL-La113 [73]. Low temperature studies of neutron diffraction and XANES across MIT show the increase of Ni-O bond length and weakening of $C$ peak intensity when setting in insulating phase [73,74]. Bond disproportionation picture has been proposed to interpret the MIT therein [65,75]. These results demonstrate the close link between NiO$_6$ octahedra distortion and conductivity via different tuning means of chemical doping, temperature, and pressure. Importantly, our XANES results unambiguously establish that the positive correlation between intensity of $C$ feature and conductivity among $RP$-nickelates, proving the pressure-induced hole doping effect.

We summarize the schematic electronic transitions of XANES for SL-La214 and BL-La327 in Fig. 6(a). All the electronic transitions are indicated for the experimental observed XANES features of $a$, $A$, $B'/B$ and $C$ peaks. Note that the minor contribution of possible $1s \rightarrow 4p3d^7$ transition for $C$ peak is not indicated. Specially, as applying pressure, sufficient hole doping would lower the $E_F$, which may cause the bonding band $d_{3z^2-r^2}$ crossing Fermi level after the SPT [15]. This has been illustrated from the integrated height of $a$ peak [Fig. 13(b)]. Two Gaussian peak are used to deconvolute the contributions of overlapped antibonding $d_{3z^2-r^2}$ and $d_{x^2-y^2}$ bands denoted as $P_L$, and the contribution of the $d_{x^2-y^2}$ bands at higher energy denoted as $P_H$.

We next examine the pressure evolution for $P \geq 7.7$ GPa, at which there is strong indication of SPT-induced change of electronic band structure topology for BL-La327-ssr sample. As mentioned before, the hole doping can be achieved when the bonding $d_{3z^2-r^2}$ band cross $E_F$ induced by pressure, forming hole-like pocket at $Q$ and $H$ point for $Fmmm$ and $X$ and $P$ point for $I4/mmm$. This would result in the growth of $P_L$. Concomitantly, the equivalent electron doping in the $d_{x^2-y^2}$ band would reduce its intensity of $P_H$ through increasing the bandwidth below $E_F$. This situation is indeed experimentally evidenced for 7.7 GPa$\leq P \leq$20 GPa [Fig. 7(b)]. Intriguingly, the $P_L$ starts to decrease but $P_H$ increase above 20 GPa, around which maximum $T_c$ reaches and higher pressure tends to suppress $T_c$ value [Fig. 7(c)]. One possibility is that new electron pockets form through lowering the antibonding $d_{3z^2-r^2}$ band at $M$ and $\Gamma$ points to cross $E_F$. It will compensate the electron doping at $d_{x^2-y^2}$ band, which in turn increases $P_H$. It is also supported by the increase of FWHM for $P_H$ above 20 GPa [Fig. 11(b)], which represents the $d_{x^2-y^2}$ bandwidth above $E_F$.

Simultaneously, FWHM of $P_L$ decreases above 20 GPa. However, other possibility of the nonmonotonic evolution of DOS due to volume contraction can't be ruled out. If the first case is at play, our results would indicate that two individual electronic transitions emerge at around 7.7 GPa and 20 GPa, which associates with the bonding and antibonding $d_{3z^2-r^2}$ bands. Notice that the $P_L$ and $P_H$ show contrasting behavior for BL-La327-sol-gel sample, i.e., the former tends to become saturated and the later one gradually declines above 20 GPa [Fig. S13(c) [55]].

Comparing with the $T_c(P)$ phase diagram [Fig. 7(c)], the evolution of area and FWHM of $P_L$ is surprisingly similar to that of $T_c(P)$ values [15-18]. Since FWHM (representing the bandwidth) actually reflects the electronic correlation strength, our results also imply that BL-La327-ssr has a strong orbital-selective electronic correlation tuned by pressure, which is negligible in BL-La327-sol-gel sample [Fig. 11(c)]. Interestingly, above 7.7 GPa, pressure tunes a $\propto$-shape of FWHM for overlapped antibonding $d_{3z^2-r^2}$ and $d_{x^2-y^2}$ bands (dominant $d_{3z^2-r^2}$), and the top $d_{x^2-y^2}$ band with the extrema emerging at about 20 GPa. Experiment and theory have shown that $T_c$ has a close link to electronic correlation strength in unconventional iron-based and cuprates superconductors [76,77], i.e., the ratio $U/t$ of on-site Coulomb repulsion $U$ and the nearest neighbor hopping magnitude $t$. To monitor the evolution of electronic correlation, we have plotted the inversed FWHM (1/FWHM) for dominant $d_{3z^2-r^2}$ and dominant $d_{x^2-y^2}$ band [Fig. 7(c)]. Clearly, orbital-selective electronic correlation emerges at about 7.7 GPa and above. Interestingly, at about 20 GPa, optimal $T_c(P)$ reaches not only at the maximum of $\Delta E$ for $3d$ bands above $E_F$, but also the maximum of electron correlation of $d_{x^2-y^2}$ band and the minimum of electronic correlation of $d_{3z^2-r^2}$ band. The balanced extrema might lead to realization of the moderate electronic correlation to optimize the $T_c$ value [76,77]. Accordingly, we suggest that the electronic correlation of both the $d_{3z^2-r^2}$ orbital and $d_{x^2-y^2}$ orbital is involved in the superconducting paring for the BL-La327 nickelate, which has been discussed by recent theories [30,33-35,37-39,42,44].

Lastly, we discuss the issue on the valence change of Ni ions induced by pressure. As discussed afore, both lattice contraction and electron screening contribute to the energy shift of main edge. Therefore, the energy shift of main edge is not a good indicator for evaluating the pressure-induced valence change of Ni ions, like the case in Sr-doped SL-La214 nickelate [58]. Instead, we adopt the energy shift of the pre-edge peak to estimate the valence change of Ni ions for it is mostly related to the $3d$ orbital states despite the existence of the hybridization with O $2p$ orbitals. Also, the edge energy just above $E_F$ associated with $d$ orbitals is almost independent with the structure factor [52], thus sensitive to the oxidation state of Ni absorber. Assuming the effective on-site Coulomb interaction $U$=3.5 eV as the energy scale for removing one electron from $Ni^{2+}$ to $Ni^{3+}$ [25], the relative energy shift 0.26 eV at 20 GPa to AP would yield a 0.074 hole per Ni site or 0.148 hole per formula unit cell (f.u.). This equivalently corresponds to $1.06 \times 10^{21}$ cm$^{-3}$ or $1.10 \times 10^{21}$ cm$^{-3}$ if taking unit cell volume $V$=556.8 Å at 19.5 GPa [15] or $V$=269.3 Å at 19.6 GPa (40 K) [22].

Nevertheless, the relative energy shift to AP becomes negative for BL-La327-sol-gel sample at HP, indicating otherwise besides hole doping. For SL-La214-ssr sample, the relative energy shift at 32.5 GPa to AP reaches 0.30 eV. This corresponds to 0.086 holes per Ni site if using the same criterion as BL-La327-ssr sample, a hole doping level far from reaching the metallicity which formally needs about 1.3 holes per f.u. by Sr doping [2]. Of course, the $U$ value may be larger than 3.5 eV for it is a CT insulator with gap $\Delta_{CT} \cong 4$ eV [5]. In SL $(La_{1-x}Sr_x)_2CuO_4$ high-$T_c$ cuprates [3], Sr doped content of only $x$=0.01 readily induces metallicity and bulk SC can be achieved at $x$=0.03 (corresponding to formal 0.06 holes per Cu sites). Therefore, our simple estimation points out that the hole doping level of pressurized BL-La327 nickelates superconductor is comparable to that of SL high-$T_c$ $(La_{1-x}Sr_x)_2CuO_4$. Further works like HP polarized XANES experiments are needed to clarify where the doped holes reside and whether they are at play in high-$T_c$ SC in pressurized BL-La327 single crystals.

## IV. CONCLUSIONS

In summary, we have reported the experimental electronic band structures of SL and BL *RP*-nickelates using a combination of *in-situ* HP XANES technique with theoretical simulations and DFT-based calculations. Main conclusions are obtained as followings.

**(i)** We have directly observed the interlayer coupling induced energy splitting of $d_{3z^2-r^2}$ and $d_{x^2-y^2}$ derived bands in BL-La327 as is absent in SL-La214 above $E_F$, which is corroborated by theoretical simulated XANES. According to the results of DFT-based band structure and PDOS, we have interpreted the $a$, $A$, $B'/B$ and $C$ features of XANES. We have quantitively clarified that the dipole transition of $1s \rightarrow 4p_{hyb}^a$ due to the hybridized Ni $3d/4p$ and O $2p$ orbitals contribute to the pre-edge $a$ peak by over 95% while the quadruple transition of $1s \rightarrow 3d$ is insignificant. The $A$ peak arises from the transition $1s \rightarrow 4p_{hyb}^A$ for the enhanced $4p$ component due to the $p-d$ hybridization of O and La atoms. The doublet main edges $B'/B$ are interpreted by the crystal field splitting and core-hole screening of different $3d$ configurations, i.e., transitions of $1s \rightarrow 4p_\pi 3d^9\underline{L}$ / $1s \rightarrow 4p_\sigma 3d^9\underline{L}$ for SL-La214, and $1s \rightarrow 4p_\pi 3d^8\underline{L}$ / $1s \rightarrow 4p_\sigma 3d^8\underline{L}$ for BL-La327. The $C$ peak is mainly caused by the transition of $1s \rightarrow 4p3d^8$ for both SL-La214 and BL-La327 nickelates.

**(ii)** We reveal that there exists contrasting evolution of the energy splitting $\Delta E$ and its deconvoluted parts of $a$ peak, in which $E_L$ represents for the overlapped $d_{3z^2-r^2}$ and $d_{x^2-y^2}$ orbitals at lower energy and $E_H$ for $d_{x^2-y^2}$ orbital at higher energy. In detail, $\Delta E$ of both the metallic and semiconducting BL-La327 samples forms a dome-like evolution with pressure for $P \geq 7.7$ GPa with the maximum locating at around 20 GPa. Nevertheless, the deconvoluted $E_L$ and $E_H$ show apparent differences. Both peak positions shift upwards and downwards to $E_F$ for 7.7 GPa$\leq P \leq 20$ GPa and $P \geq 20$ GPa for the metallic BL-La327 sample with strong superconducting transition, namely, synergetic response to pressure. However, while $E_H$ remains the upwards and downwards to $E_F$ for 4.9 GPa$\leq P \leq 15.5$ GPa and $P \geq 15.5$ GPa, $E_L$ behaves oppositely for the semiconducting BL-La327 sample with very weak superconducting transition, namely, showing opposite response to pressure. This indicates the most relevance to electrical transport properties for the overlapped

$d_{3z^2-r^2}$ and $d_{x^2-y^2}$ orbitals.

**(iii)** Growth of integrated area of the $E_L$ side and concomitant lowering of the $E_H$ side for $a$ peak provides strong evidence that the bonding $d_{3z^2-r^2}$ band crosses $E_F$ for $P \geq 7.7$ GPa. However, no such evidence is revealed in semiconducting BL-La327 sample. Intriguingly, the pressure dependent FWHM of peak evidences a nonmonotonic evolution of electronic correlation for $d_{3z^2-r^2}$ orbital and $d_{x^2-y^2}$ orbital in metallic BL-La327.

**(iv)** Pressure induced hole doping is evidenced by the growth of integrated area of $a$ peak and $C$ peak. Instead of adopting the energy shift of main edges as is mainly contributed by lattice contraction, we alternatively estimate the relative hole doping level using the energy shift of $a$ peak, yielding 0.148 hole per f.u. at 20 GPa for the metallic BL-La327 assuming $U$=3.5 eV.

Our comparative results of electronic band structures for SL-La214 and BL-La327 nickelates emphasize the essential role of both $d_{3z^2-r^2}$ and $d_{x^2-y^2}$ bands and electronic correlation in electrical transport properties, which could facilitate the understanding of the superconductivity in pressurized BL-La327.

## V. METHODS
### A. Sample synthesis, characterizations, magnetic and electrical transport measurements

Polycrystalline La$_{n+1}$Ni$_n$O$_{3n+1}$ ($n$=1 and 2) samples were prepared by solid-state reaction (ssr) method [16,78], and sol-gel method [18]. For SL-La214-ssr and BL-La327-ssr samples, stoichiometric amount of La$_2$O$_3$ (99.999%, Aladdin) and NiO (99.99%, Aladdin) were weighted and grounded thoroughly in a mortar under argon atmosphere. Then the ground mixtures were pressed into pellets and sintered at 1200-1300 °C for 48 h (SL-La214) and 1100 °C for 150 h (BL-La327) with several intermediate grounding and repressing. Note that stoichiometric La$_3$Ni$_{2.5}$O$_7$ was used to synthesize BL-La327-ssr sample, in order to compensate the oxygen content. We also synthesized BL-La327 using sol-gel mothed. Stoichiometric amount of La(NO$_3$)$_3$·6H$_2$O (AR, 99%, Aladdin) and Ni(NO$_3$)$_2$·4H$_2$O (99.9%, Aladdin) were dissolved into water, which was heated before forming a transparent, light green solution. Then, citric acid (AR, ≥99.5%, Aladdin) was added. After evaporating the water, the obtained green gel was heated to 800 °C for 6 h to produce black powders, which were pressed into dense pellets for sintering at 1150 °C for 48 h.

Powder X-ray diffraction (XRD) patterns were collected on a Bruker D2 phaser with Cu-$K_\alpha$ radiation (step <0.02° for 2theta scan) at RT. Rietveld refinement was performed by GSAS II package [79]. It is found that SL-La214-ssr and BL-La327-sol-gel are phase pure, while BL-La327-ssr has 7.9% NiO source material. Chemical composition is semi-quantitively determined by energy-dispersive X-ray spectrometry (EDS). Transport and magnetic properties were measured on the Physical Property Measurement System (Dynacool, Quantum Design) and SQUID vibrating sample magnetometer (MPMS3, Quantum Design).

HP electrical transport measurements were conducted on powder samples on a physical property measurement system (PPMS, Physike) with minimum temperature $T_{min}$ = 1.8 K. A Be-Cu alloys DAC with anvil culet sizes of 200 μm as well as BeCu gaskets were used to generate the HP non-magnetic environments in sample chamber. Mixtures of cubic boron nitrides (C-BN) powders mixed with epoxy were used for preparing the insulating layer between gasket and Pt electrodes. Electrical resistance was measured using the dc current under van der Pauw configuration [80]. Pressure was calibrated by ruby luminescence [81].

**B. DFT based calculations**

All calculations were performed in the framework of density functional theory (DFT) [82,83], using the projector augmented wave (PAW) pseudopotential method [84,85], as implemented in VASP package [86,87]. Electron exchange-correlation functional was treated in generalized gradient approximation (GGA) as proposed by Perdew-Burke-Ernzerh (PBE) [88]. Valance electrons were considered as $5s^25p^65d^16s^2$ for La, $3d^94s^1$ for Ni, and $2s^22p^4$ for O. Kinetic energy cutoff for the plane-wave basis set expansion was set to 600 eV in all the cases. Brillouin zone was sampled by a Γ-centered Monkhorst mesh with 0.2 Å$^{-1}$ as the spacing between $k$-points. Lattice parameters are fixed to the experimental refined ones [15,20,22]. Energy convergence of 1.0×10$^{-6}$ eV was used for the electronic energy minimization steps, and the force acting on each atom was set less than 0.01 eV/Å.

To correct the transition metal $d$-electron self-interaction term in the Hartree potential, a Hubbard model-like potential term was added within the Dudarev's approach [89,90], and the $U_{\text{eff}}$ for Ni was set to 4.0 eV. In order to obtain the correct orbital projection electronic structure of Ni in the local octahedral coordinate system, we rotated the lattice vector to make the direction from Ni to the interlayer O as the $z$-axis. In addition, the spin-orbit coupling (SOC) effect was taken into consideration in all the electronic structure calculations but without considering magnetism.

### C. Theoretical XAFNES simulations

Theoretical XANES simulations were performed by finite difference method near edge structure (FDMNES) code [56,57], which is mainly a fully relativistic DFT-LSDA (local spin density approximation) code allowing *ab initio* simulations [89]. In our calculations of Ni $K$-edge XANES, full-potential finite difference method (FDM) was used to solve the electronic structure. Self-consistent DFT approach was adopted with considering SOC for the heavy La atoms, which naturally make our calculations relativistic. Electronic configurations of the absorber Ni are set to be $3d^84s^2$ for initial state, and $3d^84s^24p^1$ for excited state. More details on electronic configuration setup including SOC can be found in table S2 in supplementary materials [55]. The same lattice parameters are used as DFT-based calculations that obtained from experiments at AP, i.e., $I4/mmm$ symmetry of SL-La214 [19], and for $Amam$ symmetry of BL-La327 [20]. While, HP lattice parameters of BL-La327 were adopted at 29.5 GPa for $Fmmm$ symmetry [15], 19.5 GPa $I4/mmm$ symmetry [22]. Cluster size of 7.0 Å is set to calculate the XANES ranging from -50–100 eV. Two sets of XANES with quadrupole and without quadrupole channels were calculated to evaluate the absolute contribution of quadrupole.

### D. XAFNES measurements

X-ray absorption near edge spectroscopy (XANES) measurements were conducted at I18 beamline [54], Diamond Light Source Ltd, and ODE beamline [53], synchrotron SOLEIL. At I18beamline with energy scanning application, Si(111) and Si(311) two-crystal sets are used as monochromator to allow photon energy ranging from 2 keV to 20.7 keV with energy resolution of $\Delta E/E \sim 1.4\times10^{-4}$ and $3.0\times10^{-5}$, and typical energy stability of ±0.05 eV per day. Core energy range is 5–15 keV for Si(111) reflection with high photon flux and 6–20 keV for Si(333) reflection with low flux. Mechanical Kirkpatrick–Baez (KB) mirrors with Si and Rh stripes at the end of the beamline are used to focus the beam size down to a typical 2–5 μm (Vertical) × 2–5 (Horizontal) micron spot size. Si(111) was used in our HP measurements, yielding an estimated $\Delta E/E$ ~1.17 eV at the Ni $K$-edge. Typical 0.5 eV per step was set during data acquisition at I18 beamline.

ODE is dedicated to energy-dispersive HP extended XAFS and X-ray magnetic circular dichroism (XMCD) experiments between 3.5 keV to 25 keV [53]. Si(111) was also used in our HP measurements with the same energy resolution $\Delta E/E$~$1.4\times10^{-4}$ eV but great energy stability due to the absence of any mechanical movement. Typical energy step of 0.22 eV per pixel was obtained after calibrating the XAS data collected at ODE. Typical beam spot size is about 25 μm ×35 μm in FWHM with tail of 70 μm (at 7 keV). Optical cryostat was used to conduct the low temperature (5–300 K) XAFS measurements with an open cycling liquid helium, which is transmitted by a liquid helium transfer line between He Dewar and cryostat.

Transmission mode was adopted for XAFS data acquisition at RT and low temperature. Fine powders were freshly ground from dense pellets and compressed into a pre-indented Re sample chamber without pressure transmitting medium (PTM). Membrane DAC with mini nano-polycrystalline diamonds anvils (NPDs) with thickness of 0.5 mm and 250 μm culet was used at I18 beamline [91]. Meanwhile, NPDs of culet 300 μm and 400 μm with height of 1.2-1.4 mm were used at ODE beamline. Nickel foil standard was measured for energy calibration. Gas membrane was used to realize *in-situ* pressure controlling by the PACE5000 (GE) pressure controller. Pressure was calibrated by ruby luminescence [81]. Note that, considering the BL-La327-ssr sample contain 7.9% NiO source material, we have measured several XANES for HP sample, which show generally the same features as the BL-La327-sol-gel sample, which ensures the XANES is dominantly from the BL-La327.

**ACKNOWLEDGEMENTS**

Y.P.Q. acknowledges support from the National Natural Science Foundation of China under grant No. 52272265 and the National Key R&D Program of China under grant No. 2023YFA1607400. We acknowledge the Diamond Light Source Ltd for allocating beamtime at the I18 beamline (Didcot, UK) under proposal No. SP36140, and synchrotron SOLEIL at the ODE beamline (Saint-Aubin, France) under proposal No. 20231950 and the staff of the I18 and ODE beamlines for assistance with data collection.

**AUTHOR CONTRIBUTIONS**

M.T.L. conceived and designed this project. Y.P.Q. and W.G.Y. supervised this project. M.X.Z. and Y.P.Q. synthesized and characterized the $La_2NiO_4$ and $La_3Ni_2O_7$ polycrystalline samples. P.C.Y., M.X.Z. and Y.P.Q. conducted the HP electrical transport measurements. M.A., N.-D.A., and Q.Y.K. contributed to the experimental set-up at I18 beamline, Diamond Light Source Ltd, and ODE beamline, synchrotron SOLEIL. N.N.L., M.A., J.Y.G., and M.T.L. prepared the HP gaskets and loaded the samples. M.T.L., N.N.L., J.Y.G. and W.G.Y. measured the XANES data at ambient and high pressures. M.T.L. analyzed the XANES data and performed theoretical simulations. Y.M.W. carried out the DFT-based calculations of electronic band structure and PDOS. M.T.L. wrote the manuscript with input of all authors.

**APPENDIX: Ni K-EDGE XANES, FIRST-ORDER AND SECOND ORDER DERIVATIVES, $a$, $A$, and $C$ PEAKS UNDER PRESSURE**

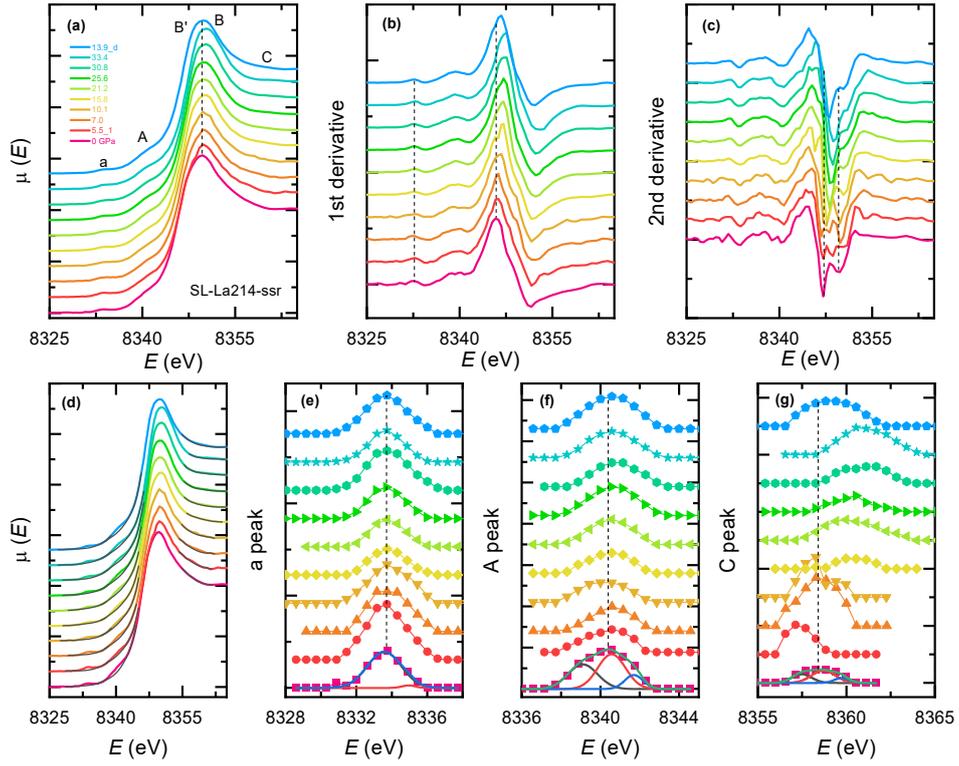

**FIG. 8. XANES, 1st derivative, 2nd derivative and subtracted *a* peak, A peak and C peak of SL-La214-ssr sample.** Dashed lines in (a)-(c) are guide to eyes. Solid lines in (d) are cubic spine lines as approximated backgrounds. Solid lines in (e)-(g) are fitting and cumulative lines by Gaussian function.

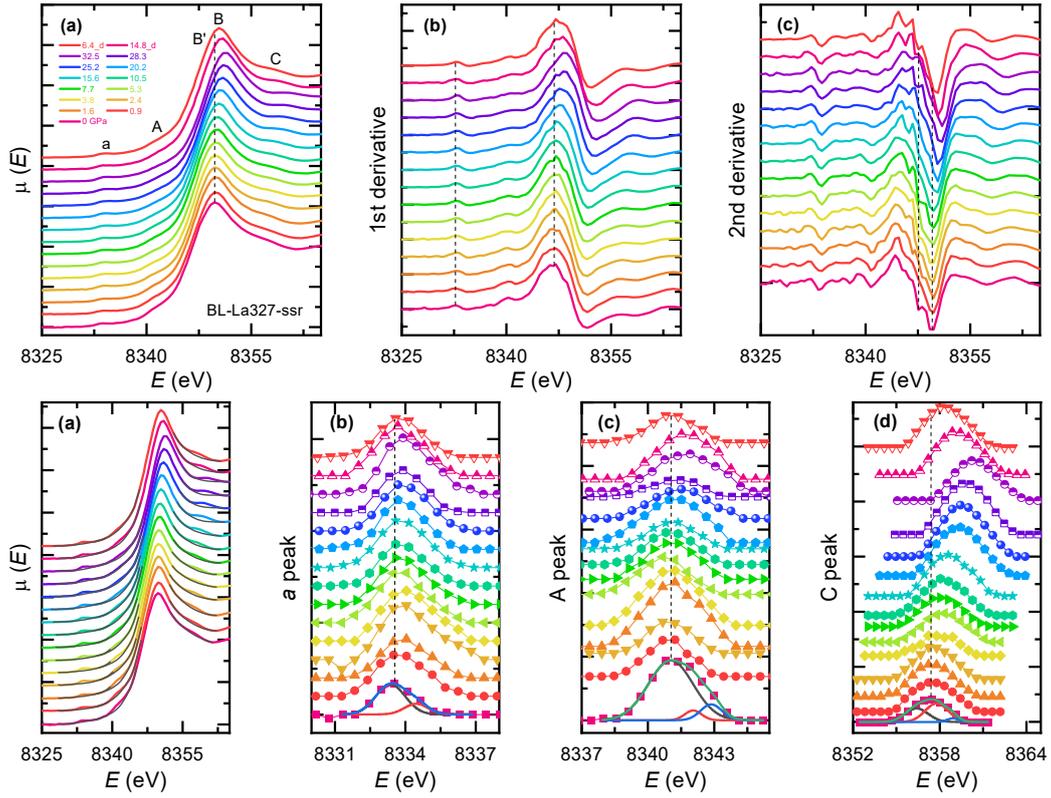

**FIG. 9. XANES, 1st derivative, 2nd derivative and subtracted *a* peak, A peak and C peak of BL-La327-ssr sample.** Dashed lines in (a)-(c) are guide to eyes. Solid lines in (d) are cubic spine lines as approximated backgrounds. Solid lines in (e)-(g) are fitting and cumulative lines by Gaussian function.

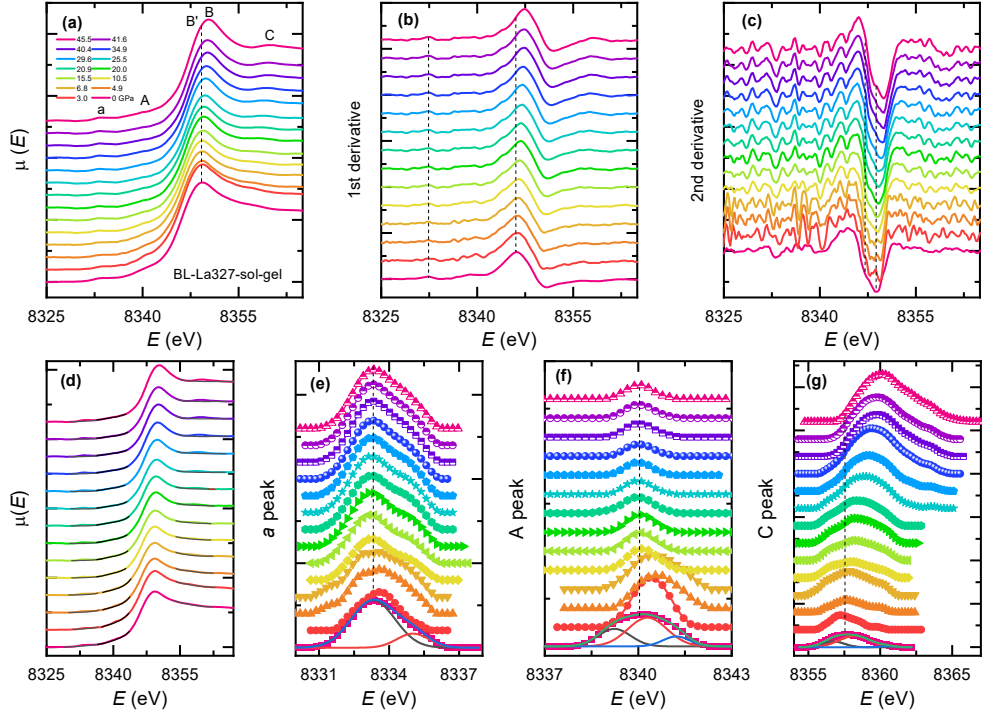

**FIG. 10. XANES, 1st derivative, 2nd derivative and subtracted *a* peak, A peak and C peak of BL-La327-sol-gel sample.** Dashed lines in (a)-(c) are guide to eyes. Solid lines in (d) are cubic spine lines as approximated backgrounds. Solid lines in (e)-(g) are fitting and cumulative lines by Gaussian function.

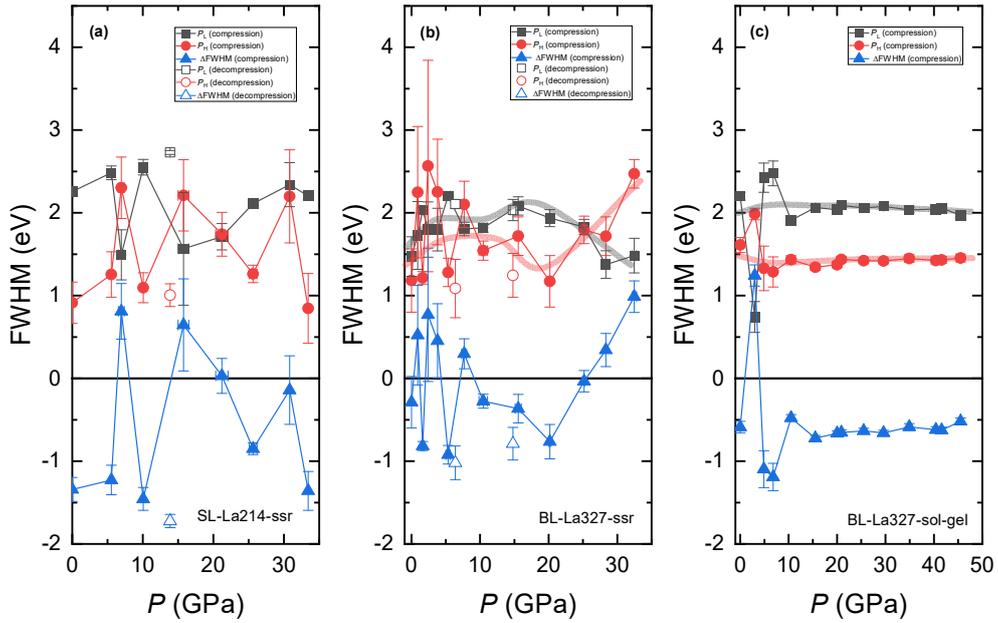

**FIG. 11. FWHM of *a* peak as a function of pressure.** (a) SL-La214-ssr. (b) BL-La327-ssr. (c) BL-La327-sol-gel. Solid bold lines are guides to eye.

**REFERENCES**

[1] J. G. Bednorz and K. A. Müller, Zeitschrift für Physik B Condensed Matter **64**, 189 (1986).

[2] Y. Takeda, R. Kanno, M. Sakano, O. Yamamoto, M. Takano, Y. Bando, H. Akinaga, K. Takita, and J. B. Goodenough, Mater. Res. Bull. **25**, 293 (1990).

[3] H. Takagi, T. Ido, S. Ishibashi, M. Uota, S. Uchida, and Y. Tokura, Physical Review B **40**, 2254 (1989).

[4] G. Aeppli and D. J. Buttrey, Physical Review Letters **61**, 203 (1988).

[5] T. Ido, K. Magoshi, H. Eisaki, and S. Uchida, Physical Review B **44**, 12094 (1991).

[6] D. Vaknin, S. K. Sinha, D. E. Moncton, D. C. Johnston, J. M. Newsam, C. R. Safinya, and H. E. King, Physical Review Letters **58**, 2802 (1987).

[7] M. T. Czyżyk and G. A. Sawatzky, Physical Review B **49**, 14211 (1994).

[8] D. Li, K. Lee, B. Y. Wang, M. Osada, S. Crossley, H. R. Lee, Y. Cui, Y. Hikita, and H. Y. Hwang, Nature **572**, 624 (2019).

[9] N. N. Wang, M. W. Yang, Z. Yang, K. Y. Chen, H. Zhang, Q. H. Zhang, Z. H. Zhu, Y. Uwatoko, L. Gu, X. L. Dong, J. P. Sun, K. J. Jin, and J. G. Cheng, Nature Communications **13**, 4367 (2022).

[10] H. Sakakibara, H. Usui, K. Suzuki, T. Kotani, H. Aoki, and K. Kuroki, Physical Review Letters **125**, 077003 (2020).

[11] A. S. Botana and M. R. Norman, Physical Review X **10**, 011024 (2020).

[12] C. Weber, C. Yee, K. Haule, and G. Kotliar, Europhysics Letters **100**, 37001 (2012).

[13] E. Pavarini, I. Dasgupta, T. Saha-Dasgupta, O. Jepsen, and O. K. Andersen, Physical Review Letters **87**, 047003 (2001).

[14] H. Sakakibara, H. Usui, K. Kuroki, R. Arita, and H. Aoki, Physical Review B **85**, 064501 (2012).

[15] H. Sun, M. Huo, X. Hu, J. Li, Z. Liu, Y. Han, L. Tang, Z. Mao, P. Yang, B. Wang, J. Cheng, D.-X. Yao, G.-M. Zhang, and M. Wang, Nature **621**, 493 (2023).

[16] M. Zhang, C. Pei, Q. Wang, Y. Zhao, C. Li, W. Cao, S. Zhu, J. Wu, and Y. Qi, Journal of Materials Science & Technology **185**, 147 (2024).

[17] Y. Zhang, D. Su, Y. Huang, Z. Shan, H. Sun, M. Huo, K. Ye, J. Zhang, Z. Yang, Y. Xu, Y. Su, R. Li, M. Smidman, M. Wang, L. Jiao, and H. Yuan, Nature Physics **20**, 1269 (2024).

[18] G. Wang, N. N. Wang, X. L. Shen, J. Hou, L. Ma, L. F. Shi, Z. A. Ren, Y. D. Gu, H. M. Ma, P. T. Yang, Z. Y. Liu, H. Z. Guo, J. P. Sun, G. M. Zhang, S. Calder, J. Q. Yan, B. S. Wang, Y. Uwatoko, and J. G. Cheng, Physical Review X **14**, 011040 (2024).

[19] S. J. Skinner, Solid State Sciences **5**, 419 (2003).

[20] C. D. Ling, D. N. Argyriou, G. Wu, and J. J. Neumeier, Journal of Solid State Chemistry **152**, 517 (2000).

[21] D. Takegami, K. Fujinuma, R. Nakamura, M. Yoshimura, K. D. Tsuei, G. Wang, N. N. Wang, J. G. Cheng, Y. Uwatoko, and T. Mizokawa, Physical Review B **109**, 125119 (2024).

[22] L. Wang, Y. Li, S.-Y. Xie, F. Liu, H. Sun, C. Huang, Y. Gao, T. Nakagawa, B. Fu, B. Dong, Z. Cao, R. Yu, S. I. Kawaguchi, H. Kadobayashi, M. Wang, C. Jin, H.-k. Mao, and H. Liu, J. Am. Chem. Soc. **146**, 7506 (2024).

[23] K. Momma and F. Izumi, Journal of Applied Crystallography **44**, 1272 (2011).

[24] X. Y. Chen, J. Choi, Z. C. Jiang, J. Mei, K. Jiang, J. Li, S. Agrestini, M. Garcia-Fernandez, X. Huang, H. L. Sun, D. W. Shen, M. Wang, J. P. Hu, Y. Lu, K. -J. Zhou, D. L. Feng, arXiv preprint **arXiv:2401.12657v1** (2024).

[25] J. Yang, H. Sun, X. Hu, Y. Xie, T. Miao, H. Luo, H. Chen, B. Liang, W. Zhu, G. Qu, C.-Q. Chen, M. Huo, Y. Huang, S. Zhang, F. Zhang, F. Yang, Z. Wang, Q. Peng, H. Mao, G. Liu, Z. Xu, T. Qian, D.-X. Yao, M. Wang, L. Zhao, and X. J. Zhou, Nature Communications **15**, 4373 (2024).

[26] Y. Li, X. Du, Y. Cao, C. Pei, M. Zhang, W. Zhao, K. Zhai, R. Xu, Z. Liu, Z. Li, J. Zhao, G. Li, Y. Qi, H. Guo, Y. Chen, and L. Yang, Chinese Physics Letters **41**, 087402 (2024).

[27] Z. Liu, M. Huo, J. Li, Q. Li, Y. Liu, Y. Dai, X. Zhou, J. Hao, Y. Lu, M. Wang, and H.-H. Wen, Nature Communications **15**, 7570 (2024).

[28] D. A. Shilenko and I. V. Leonov, Physical Review B **108**, 125105 (2023).

[29] K. Chen, X. Liu, J. Jiao, M. Zou, C. Jiang, X. Li, Y. Luo, Q. Wu, N. Zhang, Y. Guo, and L. Shu, Physical Review Letters **132**, 256503 (2024).

[30] Y. Shen, M. Qin, and G.-M. Zhang, Chinese Physics Letters **40**, 127401 (2023).

[31] Z. Luo, X. Hu, M. Wang, W. Wú, and D.-X. Yao, Physical Review Letters **131**, 126001 (2023).

[32] Y. Zhang, L.-F. Lin, A. Moreo, T. A. Maier, and E. Dagotto, Physical Review B **108**, 165141 (2023).

[33] Y.-f. Yang, G.-M. Zhang, and F.-C. Zhang, Physical Review B **108**, L201108 (2023).

[34] Q.-G. Yang, D. Wang, and Q.-H. Wang, Physical Review B **108**, L140505 (2023).

[35] H. Oh and Y.-H. Zhang, Physical Review B **108**, 174511 (2023).

[36] Z. Liao, L. Chen, G. Duan, Y. Wang, C. Liu, R. Yu, and Q. Si, Physical Review B **108**, 214522 (2023).

[37] F. Lechermann, J. Gondolf, S. Bötzel, and I. M. Eremin, Physical Review B **108**, L201121 (2023).

[38] C. Lu, Z. Pan, F. Yang, and C. Wu, Physical Review Letters **132**, 146002 (2024).

[39] H. Sakakibara, N. Kitamine, M. Ochi, and K. Kuroki, Physical Review Letters **132**, 106002 (2024).

[40] S. Ryee, N. Witt, and T. O. Wehling, Physical Review Letters **133**, 096002 (2024).

[41] Y. Zhang, L.-F. Lin, A. Moreo, T. A. Maier, and E. Dagotto, Nature Communications **15**, 2470 (2024).

[42] X.-Z. Qu, D.-W. Qu, J. Chen, C. Wu, F. Yang, W. Li, and G. Su, Physical Review Letters **132**, 036502 (2024).

[43] R. Jiang, J. Hou, Z. Fan, Z.-J. Lang, and W. Ku, Physical Review Letters **132**, 126503 (2024).

[44] J. Chen, F. Yang, and W. Li, Physical Review B **110**, L041111 (2024).

[45] Z. Ouyang, J.-M. Wang, J.-X. Wang, R.-Q. He, L. Huang, and Z.-Y. Lu, Physical

Review B **109**, 115114 (2024).

[46] Z. Dong, M. Huo, J. Li, J. Li, P. Li, H. Sun, L. Gu, Y. Lu, M. Wang, Y. Wang, and Z. Chen, Nature **630**, 847 (2024).

[47] Y. Cao and Y.-f. Yang, Physical Review B **109**, L081105 (2024).

[48] Y. Z. Zhou, J. Guo, S. Cai, H. L. Sun, P. Y. Wang, J. Y. Zhao, J. Y. Han, X. T. Chen, Y. J. Chen, Q. Wu, Y. Ding, T. Xiang, H. -K. Mao, L. L. Sun, arXiv preprint **arXiv:2311.12361v3** (2024).

[49] T. E. Westre, P. Kennepohl, J. G. DeWitt, B. Hedman, K. O. Hodgson, and E. I. Solomon, J. Am. Chem. Soc. **119**, 6297 (1997).

[50] T. Yamamoto, X-Ray Spectrometry **37**, 572 (2008).

[51] F. A. Cotton and C. J. Ballhausen, The Journal of Chemical Physics **25**, 617 (1956).

[52] C. R. Natoli, in *EXAFS and Near Edge Structure*, edited by A. Bianconi, L. Incoccia, and S. Stipcich (Springer Berlin Heidelberg, Berlin, Heidelberg, 1983), pp. 43.

[53] F. Baudelet, Q. Kong, L. Nataf, J. D. Cafun, A. Congeduti, A. Monza, S. Chagnot, and J. P. Itié, High Pressure Research **31**, 136 (2011).

[54] J. F. W. Mosselmans, P. D. Quinn, A. J. Dent, S. A. Cavill, S. D. Moreno, A. Peach, P. J. Leicester, S. J. Keylock, S. R. Gregory, K. D. Atkinson, and J. R. Rosell, Journal of Synchrotron Radiation **16**, 818 (2009).

[55] See Supplemental Material at ??? for XRD, theoretical calculations and PDOS including on-site Coulomb repulsion U, group theory and spectroscopy analysis, theoretical XANES peak fitting and energy splitting summary, experimental pre-edge peaking fitting results, evidence for lattice distortion origin of the crystal field splitting at pre-edge and main edge, and correlation between conductivity and integrated area of *C* peak under pressure.

[56] O. Bunău and Y. Joly, Journal of Physics: Condensed Matter **21**, 345501 (2009).

[57] O. Bunău, A. Y. Ramos, and Y. Joly, in *International Tables for Crystallography* (2024), pp. 752.

[58] A. Sahiner, M. Croft, S. Guha, I. Perez, Z. Zhang, M. Greenblatt, P. A. Metcalf, H. Jahns, and G. Liang, Physical Review B **51**, 5879 (1995).

[59] A. Sahiner, M. Croft, Z. Zhang, M. Greenblatt, I. Perez, P. Metcalf, H. Jhans, G. Liang, and Y. Jeon, Physical Review B **53**, 9745 (1996).

[60] N. Kosugi, H. Kondoh, H. Tajima, and H. Kuroda, Chem. Phys. **135**, 149 (1989).

[61] H. Tolentino, M. Medarde, A. Fontaine, F. Baudelet, E. Dartyge, D. Guay, and G. Tourillon, Physical Review B **45**, 8091 (1992).

[62] P. Kuiper, J. van Elp, G. A. Sawatzky, A. Fujimori, S. Hosoya, and D. M. de Leeuw, Physical Review B **44**, 4570 (1991).

[63] T. Mizokawa, H. Namatame, A. Fujimori, K. Akeyama, H. Kondoh, H. Kuroda, and N. Kosugi, Physical Review Letters **67**, 1638 (1991).

[64] V. Bisogni, S. Catalano, R. J. Green, M. Gibert, R. Scherwitzl, Y. Huang, V. N. Strocov, P. Zubko, S. Balandeh, J.-M. Triscone, G. Sawatzky, and T. Schmitt, Nature Communications **7**, 13017 (2016).

[65] R. J. Green, M. W. Haverkort, and G. A. Sawatzky, Physical Review B **94**, 195127 (2016).

[66] Y. Lu, D. Betto, K. Fürsich, H. Suzuki, H. H. Kim, G. Cristiani, G. Logvenov, N. B. Brookes, E. Benckiser, M. W. Haverkort, G. Khaliullin, M. Le Tacon, M. Minola, and B. Keimer, Physical Review X **8**, 031014 (2018).

[67] G. Bunker, in *Introduction to XAFS: A Practical Guide to X-ray Absorption Fine Structure Spectroscopy*, edited by G. Bunker (Cambridge University Press, Cambridge, 2010), pp. 106.

[68] J. Garcia, M. Benfatto, C. R. Natoli, A. Bianconi, A. Fontaine, and H. Tolentino, Chem. Phys. **132**, 295 (1989).

[69] R. Khasanov, T. J. Hicken, D. J. Gawryluk, L. P. Sorel, S. Bötzel, F. Lechermann, I. M. Eremin, H. Luetkens, Z. Guguchia, arXiv preprint **arXiv.2402.10485** (2024).

[70] L. Simonelli, V. M. Giordano, N. L. Saini, and G. Monaco, Physical Review B **84**, 195140 (2011).

[71] M. Kato, Y. Maeno, and T. Fujita, Physica C: Superconductivity **176**, 533 (1991).

[72] Z. Liu, H. Sun, M. Huo, X. Ma, Y. Ji, E. Yi, L. Li, H. Liu, J. Yu, Z. Zhang, Z. Chen, F. Liang, H. Dong, H. Guo, D. Zhong, B. Shen, S. Li, and M. Wang, Science China Physics, Mechanics & Astronomy **66**, 217411 (2022).

[73] M. Medarde, A. Fontaine, J. L. García-Muñoz, J. Rodríguez-Carvajal, M. de Santis, M. Sacchi, G. Rossi, and P. Lacorre, Physical Review B **46**, 14975 (1992).

[74] J. L. García-Muñoz, J. Rodríguez-Carvajal, P. Lacorre, and J. B. Torrance, Physical Review B **46**, 4414 (1992).

[75] T. Mizokawa, D. I. Khomskii, and G. A. Sawatzky, Physical Review B **61**, 11263 (2000).

[76] Z. R. Ye, Y. Zhang, F. Chen, M. Xu, J. Jiang, X. H. Niu, C. H. P. Wen, L. Y. Xing, X. C. Wang, C. Q. Jin, B. P. Xie, and D. L. Feng, Physical Review X **4**, 031041 (2014).

[77] M. T. Schmid, J.-B. Morée, R. Kaneko, Y. Yamaji, and M. Imada, Physical Review X **13**, 041036 (2023).

[78] G. Wu, J. J. Neumeier, and M. F. Hundley, Physical Review B **63**, 245120 (2001).

[79] B. H. Toby and R. B. Von Dreele, Journal of Applied Crystallography **46**, 544 (2013).

[80] L. J. v. d. Pauw, Philips Res. Rep **13**, 1 (1958).

[81] H. K. Mao, J. Xu, and P. M. Bell, Journal of Geophysical Research: Solid Earth **91**, 4673 (1986).

[82] P. Hohenberg and W. Kohn, Physical Review **136**, B864 (1964).

[83] W. Kohn and L. J. Sham, Physical Review **140**, A1133 (1965).

[84] P. E. Blochl, Phys Rev B Condens Matter **50**, 17953 (1994).

[85] G. Kresse and D. Joubert, Physical Review B **59**, 1758 (1999).

[86] G. Kresse and J. Furthmüller, Computational Materials Science **6**, 15 (1996).

[87] G. Kresse and J. Furthmüller, Physical Review B **54**, 11169 (1996).

[88] J. P. Perdew, K. Burke, and M. Ernzerhof, Physical Review Letters **77**, 3865 (1996).

[89] S. L. Dudarev, G. A. Botton, S. Y. Savrasov, C. J. Humphreys, and A. P. Sutton, Physical Review B **57**, 1505 (1998).

[90] M. Aykol and C. Wolverton, Physical Review B **90**, 115105 (2014).

[91] S. Petitgirard, G. Spiekermann, C. Weis, C. Sahle, C. Sternemann, and M. Wilke,

Journal of Synchrotron Radiation **24**, 276 (2017).